\documentclass[11pt,twoside, twocolumn,journal]{IEEEtran}
\usepackage[latin9]{inputenc}
\usepackage{color}
\usepackage{amsmath}
\usepackage{amsthm}
\usepackage{amssymb}
\usepackage{graphicx}

\makeatletter
\theoremstyle{plain}

\theoremstyle{plain}

\IEEEoverridecommandlockouts

\newtheorem{definition}{{Definition}}\newtheorem{theorem}{{Theorem}}\newtheorem{lemma}{{Lemma}}\newtheorem{remark}{{Remark}}\newtheorem{assumption}{{Assumption}}

\usepackage{cite}
\usepackage{amsfonts}\usepackage{algorithmic}
\usepackage{textcomp}
\usepackage{xcolor}
\def\BibTeX{{\rm B\kern-.05em{\sc i\kern-.025em b}\kern-.08em
    T\kern-.1667em\lower.7ex\hbox{E}\kern-.125emX}}

\newcommand{\black}{\color{black}}

\makeatother

\providecommand{\lemmaname}{Lemma}
\providecommand{\theoremname}{Theorem}

\begin{document}
\title{On the Rate-Distortion-Perception Function }
\author{Jun Chen, Lei Yu, Jia Wang, Wuxian Shi, Yiqun Ge, Wen Tong
}
\maketitle
\begin{abstract}
Rate-distortion-perception theory generalizes Shannon's rate-distortion
theory by introducing a constraint on the perceptual quality of the
output. The perception constraint complements the conventional distortion
constraint and aims to enforce distribution-level consistencies. In
this new theory, the information-theoretic limit is characterized
by the rate-distortion-perception function. Although  a coding theorem
for the rate-distortion-perception function has recently been established,
the fundamental nature of the optimal coding schemes remains unclear,
especially regarding the role of randomness in encoding and decoding.
It is shown in the present work that except for certain extreme cases,
the rate-distortion-perception function is achievable by deterministic
codes. This paper also clarifies the subtle differences between two
notions of perfect perceptual quality and explores some alternative
formulations of the perception constraint. 
\end{abstract}


\section{Introduction}

\label{sec:introduction}


For a Polish metric space $\mathcal{X}$, let $(\mathcal{X},\mathbb{B}(\mathcal{X}))$ be the Borel measurable space induced by the  metric.
Let $\mathcal{P}(\mathcal{X})$ denote the set of distributions (i.e., probability measures) defined over $(\mathcal{X},\mathbb{B}(\mathcal{X}))$, and let $X$ be a random variable with distribution $p_X\in\mathcal{P}(\mathcal{X})$. 
Moreover, let $\Delta:\mathcal{X}\times\mathcal{X}\rightarrow[0,\infty)$
be a  (measurable) distortion function   with  $\Delta(x,\hat{x})=0\Leftrightarrow x=\hat{x}$,  
and let $d:\mathcal{P}(\mathcal{X})\times\mathcal{P}(\mathcal{X})\rightarrow[0,\infty]$ be a divergence with $d(p_{X},p_{\hat{X}})=0\Leftrightarrow p_{X}=p_{\hat{X}}$. 
 The rate-distortion-perception function for $X$ is given by 
\begin{align}
R(D,P) & :=\inf\limits _{p_{\hat{X}|X}}I(X;\hat{X})\nonumber \\
\mbox{subject to} & \quad\mathbb{E}[\Delta(X,\hat{X})]\leq D,\label{eq:distortioncontraint}\\
 & \quad d(p_{X},p_{\hat{X}})\leq P,\label{eq:perceptionconstraint}
\end{align}
where 
the infimum is taken over all channels (i.e., Markov kernels)  $p_{\hat{X}|X}$ from $\mathcal{X}$  to itself satisfying \eqref{eq:distortioncontraint} and \eqref{eq:perceptionconstraint}.  
The rate-distortion-perception function generalizes
the classical rate-distortion function by complementing the conventional
distortion constraint (\ref{eq:distortioncontraint}) with a perception
constraint (\ref{eq:perceptionconstraint}). The rationale behind
(\ref{eq:perceptionconstraint}) is that probability distributions
have important perceptual implications, thus constraining the divergence
between $p_{X}$ and $p_{\hat{X}}$ helps enforce the perceptual consistencies
between the input and the output \cite{BM18}.

Since its inception in the seminal paper by Blau and Michaeli \cite{BM19},
rate-distortion-perception theory has received considerable attention
in the machine learning community. A coding theorem for the rate-distortion-perception
function has also been established recently \cite{TW21} (see \cite{Matsumoto18, Matsumoto19} for some variants of this coding theorem). It is worth
mentioning that there exist closely related works in the signal processing
and information theory literature on quantizer design with a prescribed
output distribution motivated by perceptual considerations \cite{LKK10, LKK11, KZLK13, SLY15J1}. This line
of research culminates in \cite{SLY15J2} with a rate-distortion theory
for output constrained lossy source coding. These two theories are
intimately connected in the sense that the perception constraint is
exactly meant to constrain the output distribution. On the other hand,
they also have noticeable differences. In particular, \cite{SLY15J2}
focuses on the case where the output is an i.i.d. sequence whereas
the formulation in \cite{BM19,TW21} does not directly impose this restriction.
It will be seen that this difference has implications in terms of
the rate-distortion tradeoff.

To understand the motivation behind the present work, it is instructive
to consider the following example first introduced in \cite{TA21}.
Let $S$ be uniformly distributed over the unit circle $\mathcal{S}:=\{s\in\mathbb{R}^{2}:\|s\|_2=1\}$, where $\|\cdot\|_p$ is the $p$-norm.
The question is how to minimize the expected distortion $\mathbb{E}[\|S-\hat{S}\|_2^{2}]$
if $S$ is encoded using $1$ bit while the reconstruction $\hat{S}$
is required to meet the perfect perceptual quality constraint (i.e.,
$\hat{S}$ is also uniformly distributed over $\mathcal{S}$). Let
$\theta(s)$ denotes the angle of $s$ for any $s\in\mathcal{S}$.
Two coding schemes are studied in \cite{TA21}. For the first coding
scheme, the encoding operation is given by 
\begin{align*}
K:=\begin{cases}
0, & \theta(S)\in[0,\pi),\\
1, & \theta(S)\in[\pi,2\pi),
\end{cases}
\end{align*}
and the decoding operation is given by 
\begin{align*}
\hat{S}:=(\cos((K+W)\pi),\sin((K+W)\pi),
\end{align*}
where $W$ is uniformly distributed over $[0,1)$ and is independent
of $S$. One can readily verify that the resulting expected distortion
\begin{align*}
\mathbb{E}[\|S-\hat{S}\|^{2}]=2-\frac{8}{\pi^{2}}.
\end{align*}
In fact, it is shown in \cite{TA21} that this is the minimum achievable
distortion with private randomness only (i.e., the random seed at
the decoder is independent of that at the encoder). The second coding
scheme makes use of $W$ at both the encoder and the decoder. Specifically,
the encoding operation is given by 
\begin{align*}
K:=\begin{cases}
0, & \frac{\theta(S)}{\pi}+W\in[0,1)\cup[2,3),\\
1, & \frac{\theta(S)}{\pi}+W\in[1,2),
\end{cases}
\end{align*}
while the decoding operation is given by 
\begin{align*}
\hat{S}:=(\cos((K-W)\pi),\sin((K-W)\pi)).
\end{align*}
In this case, we have 
\begin{align*}
\mathbb{E}[\|S-\hat{S}\|^{2}]=2-\frac{4}{\pi}<2-\frac{8}{\pi^{2}},
\end{align*}
which clearly shows the advantage of common randomness over private
randomness.

The above toy example in the one-shot setting naturally leads to the
question whether the same phenomenon appears in the asymptotic setting
where many data points are encoded at once. We shall show that the
answer depends critically on the definition of perfect perceptual
quality. Specifically, in the asymptotic setting, there are two notions
of perfect perceptual quality: weak-sense and strong-sense; the advantage
of common randomness over private randomness manifests under the strong-sense
perfect perceptual quality constraint but not under the weak-sense
version. Moreover, if the weak-sense perfect perceptual quality constraint
is further relaxed by allowing slight imperfection, then no randomness
is needed at all. We would like to point out that the difference between
common randomness and private randomness has been investigated in
the context of output constrained lossy source coding \cite{SLY15J2},
which has important implications here, especially with respect to
the case of strong-sense perfect perceptual quality.

The rest of this paper is organized as follows. Section \ref{sec:codingtheorems}
contains coding theorems for various types of coding systems; in particular,
it is shown that except for certain extreme cases, the rate-distortion-perception
function is achievable by deterministic codes. Section \ref{sec:perfectperception}
is devoted to the clarification of the subtle differences between
two different notions of perfect perceptual quality. Some alternative
formulations of the perception constraint are explored in Section
\ref{sec:interpretation}. Section \ref{sec:conclusion} concludes
the paper.

Notation:  We use $p_{X}^{n}$ to denote the product of $n$ copies of  $p_{X}$. We use $p_{Y|X}$ to denote a channel (a regular conditional distribution or a Markov kernel), which    associates to each point  $x\in \mathcal{X}$  a probability measure  $p_{Y|X}(\cdot|x)$   such that, for every measurable set $B\subseteq \mathcal{Y}$, the map $x\mapsto p_{Y|X}(B|x)$  is measurable with respect to the $\sigma$-algebra on $\mathcal{X}$. A distribution is discrete if it can be written as $\sum_i p_i \delta_{x_i}$ for some countable number of points $x_i$ and positive values $p_i$ such that $\sum_i p_i=1$, where $\delta_{x}$ is the Dirac measure at $x$. For a discrete distribution, its support is defined as the set of points at which the probability masses are  positive. 
For two distributions $p_X$ and $p_{\hat{X}}$, we use 
$\Pi(p_X,p_{\hat{X}})$ to denote the set of couplings of $p_X$ and $p_{\hat{X}}$ (i.e., the set of joint distributions $p_{X\hat{X}}$ with marginals $p_X$ and $p_{\hat{X}}$).  
Let $\mathcal{B}(\rho)$ and $\mathcal{N}(\mu,\sigma^{2})$
denote respectively the Bernoulli distribution with parameter $\rho$
and the Gaussian distribution with mean $\mu$ and variance $\sigma^{2}$.
The cardinality of set $\mathcal{S}$ is written as $|\mathcal{S}|$. 
The binary entropy function is represented by $H_{b}(\cdot)$, i.e.,
$H_{b}(a):=-a\log(a)-(1-a)\log(1-a)$ for $a\in[0,1]$. Define $1_{\mathcal{E}}(\cdot,\cdot)$ to be an indicator function in the sense that
$1_{\mathcal{E}}(x,\hat{x})=1$ if 
$(x,\hat{x}) \in \mathcal{E}$ 
and $1_{\mathcal{E}}(x,\hat{x})=0$ otherwise.  
Let $\mathrm{Unif}[i:j]$ denote the uniform distribution over $[i:j]$, where $[i:j]:=\{i,i+1,\cdots,j\}$ for integers $i\leq j$.
Throughout
this paper, the base of the logarithm function is $2$.


\section{Coding Theorems}

\label{sec:codingtheorems}

Let $\{X_{t}\}_{t=1}^{\infty}$ be an i.i.d. process with marginal
distribution $p_{X}$.

\begin{definition}\label{def:weak} 
Given distortion constraint $D$
and perception constraint $P$, rate $R$ is said to be achievable
with common randomness if for all sufficiently large $n$, there exist
shared seed distribution $p_Q$ on a Polish space $\mathcal{Q}$, encoding distribution $p_{Z|X^{n}Q}$ with  $\mathcal{Z}$      countable  (equipped  with the Hamming metric), and decoding distribution $p_{\hat{X}^{n}|ZQ}$  with  $\hat{\mathcal{X}}=\mathcal{X}$
such that the induced joint distribution $p_{X^{n}QZ\hat{X}^{n}}:=p_{X}^{n}p_{Q}p_{Z|X^{n}Q}p_{\hat{X}^{n}|ZQ}$
satisfies  
\begin{align}
 & \frac{1}{n}H(Z|Q)\leq R,\label{eq:vr}\\
 & \frac{1}{n}\sum\limits _{t=1}^{n}\mathbb{E}[\Delta(X_{t},\hat{X}_{t})]\leq D,\label{eq:D}\\
 & d(p_{X},p_{\hat{X}_{t}})\leq P,\quad t\in[1:n].\label{eq:P}
\end{align}
The infimum of such achievable $R$ is denoted by $R_{{\rm {cr}}}(D,P)$.
The achievable rate with private randomness is defined in the same
way except that the encoder and the decoder do not have access to
a shared random seed (i.e., $Q$ is set to be a constant); the corresponding
fundamental limit is denoted $R_{{\rm {pr}}}(D,P)$. If  no
randomness is allowed at all  (i.e., the encoder output $Z$  is required to be a deterministic function of $X^n$ while the decoder output $\hat{X}^n$ is required to be a deterministic function of $Z$), we denote the fundamental limit by
$R_{{\rm {nr}}}(D,P)$. 
\end{definition}

\begin{remark} The rationale behind (\ref{eq:vr}) is as follows.
Given any realization $Q=q$, random variable $Z$ can be represented
by a variable-length code of average length no more than $H(Z|Q=q)+1$.
Normalizing $H(Z|Q=q)+1$ by $n$ and taking the expectation with
respect to $Q$ yields $\frac{1}{n}H(Z|Q)+\frac{1}{n}$. The extra
factor $\frac{1}{n}$ is neglible as $n\rightarrow\infty$. Alternatively,
one can replace (\ref{eq:vr}) by $\frac{1}{n}\log|\mathcal{Z}|\leq R$,
where $\mathcal{Z}$ is the alphabet over which $Z$ is defined. This
variant is more suitable for fixed-length codes. As far
as $R_{{\rm {cr}}}(D,P)$, $R_{{\rm {pr}}}(D,P)$, and $R_{{\rm {nr}}}(D,P)$
are concerned, the difference between variable-length codes and fixed-length
codes only manifests in certain extreme cases (say, $D=0$).

\end{remark}


\begin{remark} It is easy to establish the following ordering by
invoking the operational meanings of the relevant quantities: 
\begin{align}
 & R_{{\rm {cr}}}(D,P)\leq R_{{\rm {pr}}}(D,P)\leq R_{{\rm {nr}}}(D,P).\label{eq:order1}
\end{align}
\end{remark}

In this paper, we will make some regularity assumptions along the way when they are
needed for establishing certain technical results. This first one
is as follows:

\begin{assumption}\label{assumption:convex}
	 $d(\cdot,\cdot)$ is convex in its second argument.
\end{assumption}
 
This assumption is quite mild as it is satisfied by $f$-divergence
and Rényi divergence \cite{CS04,EH14} as well as those taking the form\footnote{Such divergences arise naturally in the theory of optimal transport. In particular, 
 if $c$ is the metric on $\mathcal{X}$, then
 $d(p_X,p_{\hat{X}}):=\inf_{p_{X\hat{X}}\in\Pi(p_X,p_{\hat{X}})}\mathbb{E}[c(X,\hat{X})]$ is the $1$-Wasserstein
distance.} of
$(p_X,p_{\hat{X}}) \mapsto \inf_{p_{X\hat{X}}\in\Pi(p_X,p_{\hat{X}})}\mathbb{E}[c(X,\hat{X})]$, 
\black 
where $c$ is a 
 (measurable)  
cost function and $\Pi(p_{X},p_{\hat{X}})$ denotes the set of all couplings of
$p_{X}$ and $p_{\hat{X}}$.

The following result, due to Theis and Wagner \cite[Theorem 3]{TW21},
provides a computable characterization of $R_{{\rm {cr}}}(D,P)$ by
linking it to $R(D,P)$. \begin{theorem}\label{thm:vcr} Under Assumption \ref{assumption:convex}, 
\begin{align*}
R_{{\rm {cr}}}(D,P)=R(D,P)
\end{align*}
for $D\geq0$ and $P\geq0$. \end{theorem}

\begin{remark} It can be shown using the standard converse argument
that Theorem \ref{thm:vcr} continues to hold if (\ref{eq:P}) is
weakened to 
\begin{align*}
\frac{1}{n}\sum\limits _{t=1}^{n}d(p_{X},p_{\hat{X}_{i}})\leq P
\end{align*}
or even further weakened to 
\begin{align*}
d\left(p_{X},\frac{1}{n}\sum_{t=1}^{n}p_{\hat{X}_{t}}\right)\leq P.
\end{align*}
\end{remark}

\begin{remark} In fact, Theorem \ref{thm:vcr} is established in
\cite{TW21} under the more restrictive distortion constraint (as
compared to (\ref{eq:D})) 
\begin{align*}
\mathbb{E}[\Delta(X_{t},\hat{X}_{t})]\leq D,\quad t\in[1:n],
\end{align*}
without Assumption \ref{assumption:convex}. \end{remark}


The proof of Theorem \ref{thm:vcr} in \cite{TW21} relies on the
strong functional representation lemma \cite{LEG18} and consequently
makes use of common randomness in an essential way. Thus an open problem is posed in
 \cite{TW21}, which asks whether the same result can be established under
weaker conditions. The next result provides an affirmative answer
by showing that it suffices to use deterministic codes when $D>0$
and $P>0$. To state this result precisely, we need the following
technical assumption: 

\begin{assumption}\label{assumption:quantization}
	For any $D>0$ and $P>0$, we have
	\begin{align}
	R(D,P)<\infty; \label{eq:finiteR}
	\end{align}
	moreover, given any
	$\epsilon>0$, there exists a discrete  random
	variable $\tilde{X}$ 
	with its support $\tilde{\mathcal{X}}$ satisfying
	$\tilde{\mathcal{X}}\subseteq\mathcal{X}$ and $|\tilde{\mathcal{X}}|<\infty$
	such that 
	\begin{align}
	& I(X;\tilde{X})\leq R(D,P)+\epsilon,\label{eq:app1}\\
	& \mathbb{E}[\Delta(X,\tilde{X})]\leq D+\epsilon,\label{eq:app2}\\
	& \mathbb{E}\left[\max\limits _{\tilde{x}\in\tilde{\mathcal{X}}}\Delta(X,\tilde{x})\right]<\infty,\label{eq:app3}\\
	& d(p_{X},p_{\tilde{X}})\leq P+\epsilon,\label{eq:app4}\\
	&d(p_X,\gamma)<\infty\mbox{ for all  distributions }\gamma\mbox{  supported on }\tilde{\mathcal{X}}.\label{eq:finited}
	\end{align}
	\end{assumption}
%

Note that  (\ref{eq:finiteR}) is a prerequisite for the existence of a  coding theorem non-void for  all $D>0$ and $P>0$. It clearly holds when $|\mathcal{X}|<\infty$ since $R(D,P)\leq R(0,0)=H(X)<\infty$. When $|\mathcal{X}|<\infty$, (\ref{eq:app1})--(\ref{eq:app4}) are also trivially true. In general, by the definition of $R(D,P)$, there exists a random
variable $\hat{X}$ such that 
\begin{align*}
& I(X;\hat{X})\leq R(D,P)+\epsilon,\\
& \mathbb{E}[\Delta(X,\hat{X})]\leq D,\\
& d(p_{X},p_{\hat{X}})\leq P.
\end{align*}
If we think of $\tilde{X}$ as a quantized version of $\hat{X}$,
then (\ref{eq:app1}) is automatically satisfied due to the data processing
inequality, (\ref{eq:app3}) is to ensure that no quantization output
might be catastrophically bad while (\ref{eq:app2}) and (\ref{eq:app4})
basically require that $\mathbb{E}[\Delta(X,\hat{X})]$ and $d(p_{X},p_{\hat{X}})$
are not too sensitive to the discretization of $p_{\hat{X}}$ (which can be viewed as a form of weak convergence requirement). Finally, (\ref{eq:finited}) is  a natural consequence\footnote{Actually we only need $d(p_X,\gamma)<\infty$ for $\gamma$  in a small neighborhood of $p_{\tilde{X}}$ confined to $\mathcal{P}(\tilde{\mathcal{X}})$.} of (\ref{eq:app4}) for any reasonably behaved divergence.
So Assumption \ref{assumption:quantization} is basically always true when $|\mathcal{X}|<\infty$. 
In addition, we verify in Appendix \ref{app:assumptionquantization}  that Assumption \ref{assumption:quantization} holds for the case of square-integrable
random variable, squared distortion measure, and squared quadratic
Wasserstein distance (i.e., $\mathbb{E}[X^{2}]<\infty$, $\Delta(x,\hat{x}):=(x-\hat{x})^{2}$,
and $d(p_{X},p_{\hat{X}}):=\inf_{p_{X\hat{X}}\in\Pi(p_{X},p_{\hat{X}})}\mathbb{E}[(X-\hat{X})^{2}]$). It is worth mentioning that with deterministic
encoding and decoding performed at any finite rate, the reconstruction
is inevitably discrete. Hence, there are reasons to believe that Assumption \ref{assumption:quantization} cannot be substantially relaxed. Assumption  \ref{assumption:quantization} also has some nice implications. Specifically, together with Assumption \ref{assumption:convex}, (\ref{eq:finiteR}) implies  that $R(D,P)$ is convex and consequently  continuous in $(D,P)$ for $D>0$ and $P>0$ while (\ref{eq:finited}) implies that $d(p_{X},\gamma)$ is continuous in $\gamma$ over the interior
of the probability simplex defined on $\tilde{\mathcal{X}}$.


\begin{theorem}\label{thm:deterministic} Under Assumptions \ref{assumption:convex} and
\ref{assumption:quantization}, 
\begin{align*}
R_{{\rm {nr}}}(D,P)=R(D,P)
\end{align*}
for $D>0$ and $P>0$. \end{theorem} 
\begin{IEEEproof}
The detailed proof can be found in Appendix \ref{app:deterministic}.
The basic idea is that, in the asymtotic setting, it is possible to
leverage the aggregated randomness to simultaneously shape the marginal
distributions of all output symbols into the desired form via proper
deterministic encoding and decoding even though the bit rate might
be far below the corresponding entropy. Consider the toy example in
Section \ref{sec:introduction}. In the one-shot setting, it is clearly
impossible to simulate a uniform distribution over the unit circle
using $1$ bit if the decoder is required to be deterministic. However,
in the asymptotic setting, even if the rate remains $1$ bit per data
point, we are still able to accumulate enough randomness, which can
be shared stratigically by the reconstructed points in such a way
that they all acquire an approximate uniform distribution over the
unit circle. 
\end{IEEEproof}
The case $D=0$ corresponds to the conventional zero-error source
coding problem \cite{CT91}, for which there is no loss of optimality
in restricting the encoder and the decoder to be deterministic. Moreover,
it is clear that the perception constraint becomes superfluous when
$D=0$.
\begin{theorem}\label{thm:zeroD} For $P\geq0$, 
\begin{align*}
 & R_{{\rm {cr}}}(0,P)=R_{{\rm {pr}}}(0,P)=R_{{\rm {nr}}}(0,P)=R(0,P),
\end{align*}
where 
\begin{align*}
R(0,P)=\begin{cases}
H(X), & p_{X}\mbox{ is a discrete distribution},\\
\infty, & \mbox{otherwise}.
\end{cases}
\end{align*}
\end{theorem} 

It remains to deal with the case $P=0$. This is addressed by the
next result, for which we need the following assumption:

\begin{assumption}\label{assumption:simulation}
	For any $D>0$ and $\epsilon>0$, there exist a discrete random variable  $\tilde{X}$ and an arbitrary random variable  $\hat{X}$  on $\mathcal{X}$ such that 
  $X\leftrightarrow\tilde{X}\leftrightarrow\hat{X}$ form
	a Markov chain, the support of $\tilde{X}$, denoted $\tilde{\mathcal{X}}$,
	satisfies $|\tilde{\mathcal{X}}|<\infty$, and 
	\begin{align*}
	& I(X;\tilde{X})\leq R(D,0)+\epsilon,\\
	& \mathbb{E}[\Delta(X,\hat{X})]\leq D,\\
	& p_{\hat{X}}=p_{X}.
	\end{align*}
	\end{assumption}
 
Note that according to the definition of $R(D,0)$, there exists a random variable $\hat{X}$ such that 
\begin{align*}
 & I(X;\hat{X})\leq R(D,0)+\epsilon,\\
& \mathbb{E}[\Delta(X,\hat{X})]\leq D,\\
& p_{\hat{X}}=p_{X}.
\end{align*}
So Assumption \ref{assumption:simulation} basically postulates the existence of a discrete random variable $\tilde{X}$ sitting between $X$ and $\hat{X}$ with $I(X;\tilde{X})\approx I(X;\hat{X})$. This is trivially true when $|\mathcal{X}|<\infty$. In general, this assumption is quite natural as even with the availability of private randomness, the interface between the encoder and the decoder remains discrete at any finite rate. We verify at the end of Appendix \ref{app:verification} that Assumption \ref{assumption:simulation} holds  for the case of square-integrable
random variable and squared distortion measure.

\begin{theorem}\label{thm:weakP} Under Assumptions \ref{assumption:convex} and \ref{assumption:simulation}, 
\begin{align*}
R_{{\rm {pr}}}(D,0)=R(D,0)
\end{align*}
for $D>0$. \end{theorem} \begin{remark} It is clear that deterministic
encoder-decoder pairs are  inadequate for achieving finite-valued $R(D,0)$
when  $p_{X}$ is a continuous distribution or has an infinite entropy. \end{remark} 
\begin{IEEEproof}
See Appendix \ref{app:weakP}. 
\end{IEEEproof}


\section{On Different Notions of Perfect Perceptual Quality}

\label{sec:perfectperception}


Note that setting $P=0$ in (\ref{eq:P}) only ensures $p_{\hat{X}_{t}}=p_{X_{t}}$,
$t\in[1:n]$, which should be distinguished from the more restrictive
constraint $p_{\hat{X}^{n}}=p_{X}^{n}$. We shall refer to the former
as weak-sense perfect perceptual quality and the latter as strong-sense
perfect perceptual quality. It is interesting to understand whether
these two notions of perfect perceptual quality make any difference
in terms of the rate-distortion tradeoff.



Let $R_{{\rm {cr}}}(D):=R_{{\rm {cr}}}(D,0)$ and $R_{{\rm {pr}}}(D):=R_{{\rm {pr}}}(D,0)$.
In light of Theorems \ref{thm:vcr}, \ref{thm:zeroD}, and \ref{thm:weakP},
for $D\geq0$, 
\begin{align}
R_{{\rm {cr}}}(D)=R_{{\rm {pr}}}(D)=\phi(D),\label{eq:weakperception}
\end{align}
where 
\begin{align*}
\phi(D):=R(D,0) & =\inf\limits _{p_{\hat{X}|X}}I(X;\hat{X})\\
\mbox{subject to} & \quad\mathbb{E}[\Delta(X,\hat{X})]\leq D,\\
 & \quad p_{\hat{X}}=p_{X}.
\end{align*}

Now we proceed to define the counterparts of $R_{{\rm {cr}}}(D)$
and $R_{{\rm {pr}}}(D)$ under the strong-sense perfect perceptual
quality constraint.

\begin{definition} Given distortion constraint $D$, rate $\tilde{R}$
is said to be achievable with common randomness under the strong-sense
perfect perceptual quality constraint if for all sufficiently large
$n$, there exist shared seed distribution $p_Q$ (on a Polish space), encoding distribution $p_{Z|X^{n}Q}$ (with $\mathcal{Z}$ countable), and
decoding distribution $p_{\hat{X}^{n}|ZQ}$  (with $\hat{\mathcal{X}}=\mathcal{X}$) such that the induced joint distribution
$p_{X^{n}QZ\hat{X}^{n}}:=p_{X}^{n}p_{Q}p_{\hat{X}^{n}|ZQ}p_{\hat{X}^{n}|ZQ}$
satisfies 
\begin{align*}
 & \frac{1}{n}H(Z|Q)\leq\tilde{R},\\
 & \frac{1}{n}\sum\limits _{t=1}^{n}\mathbb{E}[\Delta(X_{t},\hat{X}_{t})]\leq D,\\
 & p_{\hat{X}^{n}}=p_{X}^{n}.
\end{align*}
The infimum of such achievable $\tilde{R}$ is denoted $\tilde{R}_{{\rm {cr}}}(D)$.
In the case of private randomness only, the corresponding limit is
denoted $\tilde{R}_{{\rm {pr}}}(D)$. 
\end{definition} \begin{remark} Analogously to (\ref{eq:order1}),
we have 
\begin{align}
 & \tilde{R}_{{\rm {cr}}}(D)\leq\tilde{R}_{{\rm {pr}}}(D).\label{eq:crpr}
\end{align}
Moreover, since strong-sense perfect perceptual quality implies weak-sense
perfect perceptual quality, it follows that 
\begin{align*}
R_{{\rm {cr}}}(D)\leq\tilde{R}_{{\rm {cr}}}(D),\quad R_{{\rm {pr}}}(D)\leq\tilde{R}_{{\rm {pr}}}(D).
\end{align*}
\end{remark} \begin{remark} Under the strong-sense perfect perceptual
quality constraint, requiring the encoder and the decoder to be deterministic
trivializes the problem as the encoder-decoder pair is basically forced
to establish a one-to-one mapping between the input and the output.
\end{remark} 

The following result, together with (\ref{eq:weakperception}), shows
that in the presence of common randomness, the difference between
weak-sense perfect perceptual quality and strong-sense perfect perceptual
quality has no impact on the fundamental rate-distortion tradeoff.
\begin{theorem}\label{thm:crstrong} For $D\geq0$ 
\begin{align}
R_{{\rm {cr}}}(D)=\tilde{R}_{{\rm {cr}}}(D)=\phi(D).\label{eq:strongcr}
\end{align}
\end{theorem} 
\begin{IEEEproof}
This result can be deduced from the proof of \cite[Theorem 7]{SLY15J1}
(see also \cite{LKK11}). 
\end{IEEEproof}
The following result, together with (\ref{eq:weakperception}), shows
that in the case of private randomness only, the two different notions
of perfect randomness indeed lead to different rate-distortion tradeoffs.
Along with Theorem \ref{thm:crstrong}, it also indicates that under
the strong-sense perfect perceptual quality constraint, common randomness
is generally more powerful than private randomness, which should be
contrasted with the fact that under the weak-sense perfect perceptual
quality constraint, the difference between common randomness and private
randomness is immaterial (see Theorem \ref{thm:weakP}).

 We first introduce a definition, which is an extended version of \cite[Definition 3]{Wagner22}.
\begin{definition}\label{def:integrable}
	A tuple $(p_{X},p_{\hat{X}},\Delta)$
	of source distribution, reconstruction distribution, and distortion
	measure is said to be uniformly integrable if for every $\epsilon>0$, there
	exists $\delta>0$ such that $\sup_{p_{X\hat{X}},\mathcal{E}}\mathbb{E}[\Delta(X,\hat{X})1_{\mathcal{E}}(X,\hat{X})]\le\epsilon$,
	where the supremum is over all $p_{X\hat{X}}\in\Pi(p_X,p_{\hat{X}})$
	and
	all measurable events $\mathcal{E}$ with $\mathbb{P}((X,\hat{X})\in\mathcal{E})\le\delta$.
\end{definition}

Note that $(p_{X},p_{\hat{X}},\Delta)$ is uniformly integrable if $\Delta$ is bounded, i.e.,  $\sup_{x,\hat{x}\in\mathcal{X}}\Delta(x,\hat{x})<\infty$, which is trivially true when $|\mathcal{X}|<\infty$. Moreover, in Appendix \ref{app:uniformintegrability}, we verify uniform integrability  for square-integrable $X$ and $\hat{X}$ paired with squared distortion measure.

\begin{figure}[htbp]
\centerline{\includegraphics[width=9cm]{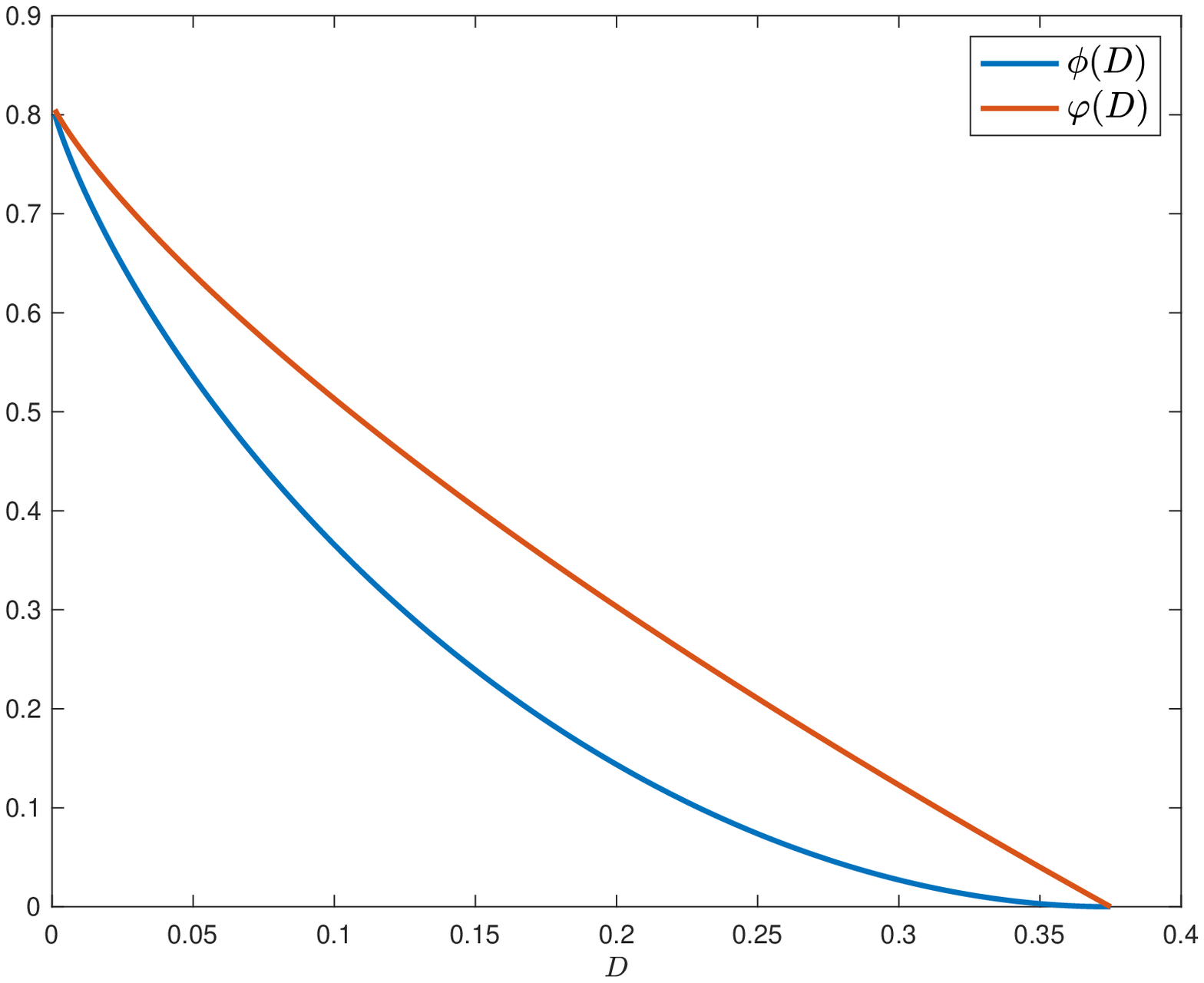}} \caption{Plots of $\phi(D)$ and $\varphi(D)$ for $X\sim\mathcal{B}(\frac{1}{4})$.}
\label{fig:binary} 
\end{figure}

\begin{theorem}\label{thm:prstrong} 
If $(p_X,p_X,\Delta)$ is uniformly integrable, then
\begin{align}
\tilde{R}_{{\rm {pr}}}(D)=\varphi(D)\label{eq:prstrong}
\end{align}
for $D\geq0$, where\footnote{
Here $\hat{\mathcal{X}}=\mathcal{X}$, and the infimum above is taken over all $p_{U\hat{X}|X}$ with $\mathcal{U}$ being a Polish space such that \eqref{eq:constraint1}-\eqref{eq:constraint3} hold. A similar convention applies to the bound in Theorem \ref{thm:prtensorization}.}
\begin{align}
\varphi(D) & :=\inf\limits _{p_{U\hat{X}|X}}\max\{I(X;U),I(\hat{X};U)\}\nonumber \\
\mbox{subject to} & \quad\mathbb{E}[\Delta(X,\hat{X})]\leq D,\label{eq:constraint1}\\
 & \quad p_{\hat{X}U|X}=p_{U|X}p_{\hat{X}|U},\label{eq:constraint2}\\
 & \quad p_{\hat{X}}=p_{X}.\label{eq:constraint3}
\end{align}
Moreover, under the squared distortion measure (assuming $\mathcal{X}\subseteq\mathbb{R}$), 
\begin{align}
\varphi(D)=R(\frac{D}{2}),\label{eq:double}
\end{align}
where 
\begin{align}
R(\frac{D}{2}) & :=\inf_{p_{V|X}}I(X;V)\label{eq:optimization}\\
\mbox{subject to} & \quad\mathbb{E}[(X-V)^{2}]\leq\frac{D}{2}.\label{eq:constraintV}
\end{align}
\end{theorem} \begin{remark} $R(\frac{D}{2})$ is interpreted as
$R(\frac{D}{2},\infty)$ in \cite{BM19,YWYML21}. This interpretation
is actually not completely accurate. Note that $R(D)$ is the rate-distortion
function with the output alphabet being $\mathbb{R}$ as $V$ is allowed
to take any real value. In constrast, $R(D,\infty)$ is the rate-distortion
function with the output alphabet being $\mathcal{X}$. In general,
under the squared distortion measure, 
\begin{align*}
R(D,0)\leq R(\frac{D}{2})\leq R(\frac{D}{2},\infty),
\end{align*}
where the second inequality becomes an equality when $\mathcal{X}=\mathbb{R}$.
Note that the first inequality follows by (\ref{eq:crpr}), (\ref{eq:strongcr}),
(\ref{eq:prstrong}), and (\ref{eq:double}) (see also \cite[Theorem 2]{BM19}).
\end{remark} 
\begin{IEEEproof}
One can specialize (\ref{eq:prstrong}) from \cite[Theorem 1]{SLY15J2} and \cite[Theorem 2]{Wagner22}.
 Moreover, (\ref{eq:double}) is implied by \cite[Theorem 2]{YWYML21}
(see also \cite[Theorem]{TA21}). We give a simple proof of this fact
in Appendix \ref{app:prstrong}. 
\end{IEEEproof}
\begin{theorem}\label{thm:binary} For $X\sim\mathcal{B}(\rho)$
with $\rho\in(0,\frac{1}{2}]$, 
\begin{align*}
 & \phi(D)=\begin{cases}
2H_{b}(\rho)+\frac{2-2\rho-D}{2}\log(\frac{2-2\rho-D}{2})+D\log(\frac{D}{2})\\
+\frac{2\rho-D}{2}\log(\frac{2\rho-D}{2}), & \hspace{-1.25in}D\in[0,2\rho(1-\rho)),\\
0, & \hspace{-1.25in}D\in[2\rho(1-\rho),\infty),
\end{cases}\\
 & \varphi(D)=\begin{cases}
H_{b}(\rho)-H_{b}(\frac{1-\sqrt{1-2D}}{2}), & D\in[0,2\rho(1-\rho)),\\
0, & D\in[2\rho(1-\rho),\infty),
\end{cases}
\end{align*}
under the Hamming distortion measure (which coincides with the squared
distortion measure when $\mathcal{X}=\{0,1\}$). \end{theorem} \begin{remark}
In this case, we can deduce from \cite[Equation (6)]{BM19} that 
\begin{align*}
R(\frac{D}{2},\infty)=\begin{cases}
H_{b}(\rho)-H_{b}(\frac{D}{2}), & D\in[0,2\rho),\\
0, & D\in[2\rho,\infty),
\end{cases}
\end{align*}
which is in general different from $\varphi(D)$. So \cite[Theorem 2]{YWYML21}
should be interpreted with great caution.

For illustrative purposes, we plot $\phi(D)$ and $\varphi(D)$ for
$X\sim\mathcal{B}(\frac{1}{4})$ in Fig. \ref{fig:binary}. \end{remark}

\begin{IEEEproof}
The expression of $\phi(D)$ can be specialized from \cite[Equation (6)]{BM19}.
The derivation of $\varphi(D)$ is given in Appendix \ref{app:binary}. 
\end{IEEEproof}

\begin{figure}[htbp]
\centerline{\includegraphics[width=9cm]{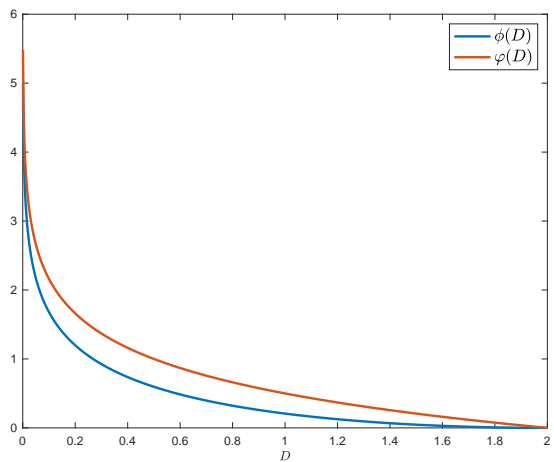}} \caption{Plots of $\phi(D)$ and $\varphi(D)$ for $X\sim\mathcal{N}(0,1)$.}
\label{fig:Gaussian} 
\end{figure}

\begin{theorem}\label{thm:Gaussian} For $X\sim\mathcal{N}(\mu,\sigma^{2})$,
\begin{align*}
 & \phi(D)=\begin{cases}
\frac{1}{2}\log(\frac{4\sigma^{4}}{4\sigma^{2}D-D^{2}}), & D\in[0,2\sigma^{2}),\\
0, & D\in[2\sigma^{2},\infty),
\end{cases}\\
 & \varphi(D)=\begin{cases}
\frac{1}{2}\log(\frac{2\sigma^{2}}{D}), & D\in[0,2\sigma^{2}),\\
0, & D\in[2\sigma^{2},\infty),
\end{cases}
\end{align*}
under the squared distortion measure. \end{theorem} \begin{remark}
In this case, $\mathcal{X}=\mathbb{R}$ and consequently \cite[Theorem 2]{YWYML21}
holds. Indeed, we can deduce from \cite[Theorem 1]{ZQCK21} that 
\begin{align*}
R(\frac{D}{2},\infty)=\begin{cases}
\frac{1}{2}\log(\frac{2\sigma^{2}}{D}), & D\in[0,2\sigma^{2}),\\
0, & D\in[2\sigma^{2},\infty),
\end{cases}
\end{align*}
which coincides with $\varphi(D)$.

For illustrative purposes, we plot $\phi(D)$ and $\varphi(D)$ for
$X\sim\mathcal{N}(0,1)$ in Fig. \ref{fig:Gaussian}. \end{remark}

\begin{IEEEproof}
The expression of $\phi(D)$ can be specialized from \cite[Theorem 1]{ZQCK21} 
(see also \cite[Proposition 2]{LKK11}) while the expression of $\varphi(D)$
can be obtained by invoking (\ref{eq:double}) and the fact \cite{CT91}
that for the quadratic Gaussian case, 
\begin{align*}
 & R(D)=\begin{cases}
\frac{1}{2}\log(\frac{\sigma^{2}}{D}), & D\in[0,\sigma^{2}),\\
0, & D\in[2\sigma^{2},\infty).
\end{cases}
\end{align*}
Various extensions can be found in \cite{LZCK22, Wagner22}.
\end{IEEEproof}








\section{Alternative Formulations of the Perception Constraint}

\label{sec:interpretation}

The fact that the perception constraint in (\ref{eq:P}) fails to
capture the notion of strong-sense perpect perceptual quality motivates
us to consider the following alternative formulation.

\begin{definition}\label{def:strong} Given distortion constraint
$D$ and perception constraint $P$, rate $\tilde{R}$ is said to
be achievable with common randomness if for all sufficiently large
$n$, there exist shared seed distribution $p_Q$ (on a Polish space), encoding distribution $p_{Z|X^{n}Q}$ (with $\mathcal{Z}$ countable), and
decoding distribution $p_{\hat{X}^{n}|ZQ}$  (with $\hat{\mathcal{X}}=\mathcal{X}$) such that the induced joint distribution
$p_{X^{n}QZ\hat{X}^{n}}:=p_{X}^{n}p_{Q}p_{Z|X^{n}Q}p_{\hat{X}^{n}|ZQ}$
satisfies 
\begin{align}
 & \frac{1}{n}H(Z|Q)\leq\tilde{R},\nonumber \\
 & \frac{1}{n}\sum\limits _{t=1}^{n}\mathbb{E}[\Delta(X_{t},\hat{X}_{t})]\leq D,\nonumber \\
 & \frac{1}{n}d(p_{X}^{n},p_{\hat{X}^{n}})\leq P.\label{eq:Pstrong}
\end{align}
The infimum of such achievable $\tilde{R}$ is denoted $\tilde{R}_{{\rm {cr}}}(D,P)$.
In the case of private randomness only, the corresponding limit is
denoted $\tilde{R}_{{\rm {pr}}}(D,P)$. \end{definition}

Note that the strong-sense perceptual quality constraint can be viewed
as an extreme case of the new formulation since setting $P=0$ in
(\ref{eq:Pstrong}) forces $p_{\hat{X}^{n}}=p_{X}^{n}$. So we have
$\tilde{R}_{{\rm {cr}}}(D,0)=\tilde{R}_{{\rm {cr}}}(D)$ and $\tilde{R}_{{\rm {pr}}}(D,0)=\tilde{R}_{{\rm {pr}}}(D)$.
However, the multiletter nature of the perception constraint in (\ref{eq:Pstrong})
makes it challenging to obtain a computable characterization of $\tilde{R}_{{\rm {cr}}}(D,P)$
and $\tilde{R}_{{\rm {pr}}}(D,P)$ when $P>0$. To ease the difficulty,
we shall impose some further restrictions on $d(\cdot,\cdot)$: 

\begin{assumption}\label{assumption:tensorizable}
	$d(\cdot,\cdot)$ is tensorizable in the sense that 
	\begin{align*}
	d(\otimes_{t=1}^{n}p_{Y_{t}},p_{\hat{Y}^{n}})\geq\sum\limits _{t=1}^{n}d(p_{Y_{t}},p_{\hat{Y}_{t}}),
	\end{align*}
	where $\otimes_{t=1}^{n}p_{Y_{i}}$ denotes the joint distribution
	formed by the product of marginals $p_{Y_{1}},\cdots,p_{Y_{n}}$.
\end{assumption}

\begin{assumption}\label{assumption:decomposable}
	$d(\cdot,\cdot)$ is decomposable in the sense that 
	\begin{align*}
	d(\otimes_{t=1}^{n}p_{Y_{t}},\otimes_{t=1}^{n}p_{\hat{Y}_{t}})=\sum\limits _{t=1}^{n}d(p_{Y_{t}},p_{\hat{Y}_{t}}).
	\end{align*}
\end{assumption}

Note that Assumptions \ref{assumption:tensorizable} and \ref{assumption:decomposable} are satisfied by the Kullback-Leibler divergence and those taking the form of $\inf_{p_{Y^n\hat{Y}^n}\in\Pi(p_{Y^n},p_{\hat{Y}^n})}\mathbb{E}[c(Y^n,\hat{Y}^n)]$ with an additive cost function $c$ in the sense $c(y^n,\hat{y}^n)=\sum_{i=1}^nc'(y_i,\hat{y}_i)$ for some $c'$ (e.g., $c(y^n,\hat{y}^n):=\|y^n-\hat{y}^n\|^p_p$).




\begin{theorem}\label{thm:crtensorization} Under Assumptions \ref{assumption:convex},
\ref{assumption:tensorizable}, and \ref{assumption:decomposable}, 
\begin{align*}
\tilde{R}_{{\rm {cr}}}(D,P)=R(D,P)
\end{align*}
for $D\geq0$ and $P\geq0$. \end{theorem} 
\begin{IEEEproof}
We first prove the converse part. Let $\tilde{R}$ be an achievable
rate with respect to distortion constraint $D$ and perception constraint
$P$. We have 
\begin{align}
\tilde{R} & \geq\frac{1}{n}H(Z|Q)\nonumber \\
 & \geq\frac{1}{n}I(X^{n};Z|Q)\nonumber \\
 & \geq\frac{1}{n}I(X^{n};\hat{X}^{n}|Q)\nonumber \\
 & =\frac{1}{n}\sum\limits _{t=1}^{n}I(X_{t};\hat{X}^{n}|Q,X^{t-1})\nonumber \\
 & =\frac{1}{n}\sum\limits _{t=1}^{n}I(X_{t};\hat{X}^{n},Q,X^{t-1})\nonumber \\
 & \geq\frac{1}{n}\sum\limits _{t=1}^{n}I(X_{t};\hat{X}_{t})\nonumber \\
 & =I(X_{T};\hat{X}_{T}|T)\nonumber \\
 & =I(X_{T};\hat{X}_{T},T)\nonumber \\
 & \geq I(X_{T};\hat{X}_{T}),\label{eq:comb1}
\end{align}
where $T$ is uniformly distributed over $[1:n]$ and independent
of $(X^{n},\hat{X}^{n})$. Moreover, 
\begin{align}
D & \geq\frac{1}{n}\sum\limits _{t=1}^{n}\mathbb{E}[\Delta(X_{t},\hat{X}_{t})]\nonumber \\
 & =\mathbb{E}[\mathbb{E}[\Delta(X_{T},\hat{X}_{T})|T]]\nonumber \\
 & =\mathbb{E}[\Delta(X_{T},\hat{X}_{T})],\label{eq:comb2}
\end{align}
and 
\begin{align}
P & \geq\frac{1}{n}d(p_{X}^{n},p_{\hat{X}^{n}})\nonumber \\
 & \geq\frac{1}{n}\sum\limits _{t=1}^{n}d(p_{X_{t}},p_{\hat{X}_{t}})\label{eq:tensor}\\
 & =\frac{1}{n}\sum\limits _{t=1}^{n}d(p_{X},p_{\hat{X}_{t}})\nonumber \\
 & \geq d\left(p_{X},\frac{1}{n}\sum\limits _{t=1}^{n}p_{\hat{X}_{t}}\right)\label{eq:convex}\\
 & =d(p_{X},p_{\hat{X}_{T}}),\label{eq:comb3}
\end{align}
where (\ref{eq:tensor}) and (\ref{eq:convex}) are due to respectively
the tensorizability and convexity (in its second argument) of $d(\cdot,\cdot)$.
Combining (\ref{eq:comb1}), (\ref{eq:comb2}), (\ref{eq:comb3})
and invoking the fact that $p_{X_{T}}=p_{X}$ proves $\tilde{R}_{{\rm {cr}}}(D,P)\geq R(D,P)$.

Now it remains to prove the achievability part. We augment $\{X_{t}\}_{t=1}^{\infty}$
into a joint i.i.d. process $\{(X_{t},\hat{X}_{t})\}_{t=1}^{\infty}$
using a memoryless channel $p_{\hat{X}|X}$ satisfying $\mathbb{E}[\Delta(X,\hat{X})]\leq D$
and $d(p_{X},p_{\hat{X}})\leq P$. For any positive integer $n$,
\begin{align*}
\frac{1}{n}d(p_{X}^{n},p_{\hat{X}^{n}})=d(p_{X},p_{\hat{X}})\leq P,
\end{align*}
where the equality follows by the decomposability of $d(\cdot,\cdot)$.
Moreover, by the strong functional representation lemma \cite{LEG18},
there exists a random variable $Q$, independent of $X^{n}$, such
that $\hat{X}^{n}$ can be expressed as a deterministic function of
$(X^{n},Q)$ and 
\begin{align*}
\frac{1}{n}H(\hat{X}^{n}|Q)\leq I(X;\hat{X})+\frac{1}{n}\log(nI(X;\hat{X})+1)+\frac{4}{n}.
\end{align*}
Setting $Z=\hat{X}^{n}$ and sending $n\rightarrow\infty$ completes
the proof. 
\end{IEEEproof}

\begin{theorem}\label{thm:prtensorization} 
For $D\geq 0$ and $P\geq 0$,
\begin{align*}
\tilde{R}_{{\rm {pr}}}(D,P)\leq R'(D,P),
\end{align*}
where 
\begin{align*}
R'(D,P) & :=\inf\limits _{p_{U\hat{X}|X}}\max\{I(X;U),I(\hat{X};U)\}\\
\mbox{subject to} & \quad\mathbb{E}[\Delta(X,\hat{X})]\leq D,\\
 & \quad p_{\hat{X}U|X}=p_{U|X}p_{\hat{X}|U},\\
 & \quad d(p_{X},p_{\hat{X}})\leq P,\\
 & \quad (p_X,p_{\hat{X}},\Delta)\mbox{ is uniformly integrable}.
\end{align*}
\end{theorem} 
\begin{IEEEproof}
Suppose that $\hat{X}^{n}$ is constrained to be an i.i.d. sequence
with marginal distribution $p_{\hat{X}}$ satisfying $d(p_{X},p_{\hat{X}})\leq P$.
For such $\hat{X}^{n}$, 
\begin{align*}
\frac{1}{n}d(p_{X}^{n},p_{\hat{X}^{n}})=d(p_{X},p_{\hat{X}})\leq P,
\end{align*}
where the equality follows by the decomposability of $d(\cdot,\cdot)$.
Now the problem is converted to output constrained lossy source coding
in the sense of \cite{SLY15J2}, and consequently the desired upper
bound can be deduced by following steps similar to those for the achievability part of \cite[Theorem 2]{Wagner22}.

It can be shown that, with the output constrained to be i.i.d., this
upper bound is in fact the best possible (assuming $d(\cdot,\cdot)$
is tensorizable, decomposable, and convex in its second argument).
Moreover, when $P=0$, the i.i.d. constraint is automatically satisfied
and the upper bound is tight. Indeed, we have $R'(D,0)=\varphi(D)$,
which is known to coincide with $\tilde{R}_{{\rm {pr}}}(D,0)$ (or
equivalently, $\tilde{R}_{{\rm {pr}}}(D)$) according to Theorem \ref{thm:prstrong}.
On the other hand, it is unclear whether the i.i.d. constraint is
redundant when $P>0$. So the upper bound can potentially be loose
in that regime. In constrast, as the proof of Theorem \ref{thm:crtensorization}
indicates, with the presense of common randomness, the i.i.d. constraint
incurs no penalty in terms of the rate-distortion-perception tradeoff. 
\end{IEEEproof}
Roughly speaking, distortion measures and perception measures in the
rate-distortion-perception framework can be considered full-reference
metrics and no-reference metrics, respectively. Take image compression
as an example. To evaluate a compressed image, full-reference metrics
compare it to the ground truth (i.e., the original version) while
no-reference metrics quantify its quality using a prescribed criterion
based on some statistical feature information. In the current setting,
the realization of $\hat{X}^{n}$, the realization of $X^{n}$, and
the distribution of $X^{n}$ correspond to the object to be evaluated,
the ground truth, and the statistical feature information, respectively.
However, the perception constraint in the form of (\ref{eq:P}) or
(\ref{eq:Pstrong}) is imposed on the distribution of $\hat{X}^{n}$,
not on its realization. This is somewhat unsatisfactory since the
perceptual quality should be defined for each individual object (say,
an image) rather than for an ensemble only. Furthermore, it is often
impossible to deduce a realization-based perceptual quality measure
from a distribution-based measure because two different distributions
may generate the same realization. So the operational meaning of (\ref{eq:P})
and (\ref{eq:Pstrong}) is not entirely clear.

\begin{figure}[htbp]
\centerline{\includegraphics[width=9cm]{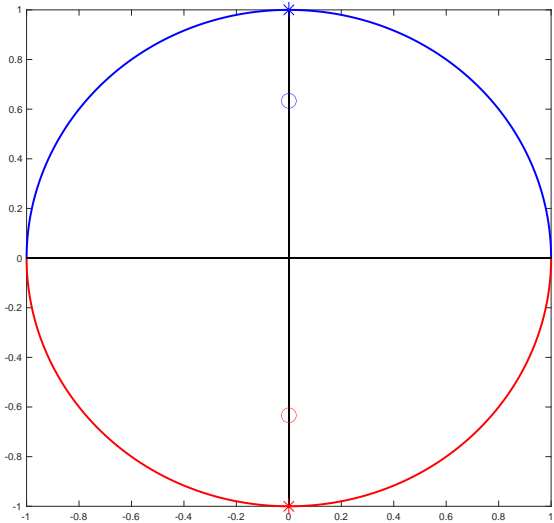}} \caption{A toy example. The reconstruction points under the perception constraint
are highlighted with $*$ while the unconstrained counterparts are
highlighted with $\circ$.}
\label{fig:circle} 
\end{figure}

To gain a better understanding, let us revisit the toy example in
Section \ref{sec:introduction}. Suppose we want to evaluate the ``perceptual
quality" of a given realization of $\hat{S}$. Clearly, it is irrelevant
here whether the distribution on the unit circle is uniform or not
as we are dealing with a property of the realization itself. When
a particular realization is concerned, the only role of the so-called
``perfect perceptual quality constraint" is to force it to be on
the unit circle. So the ``perception constraint" should be better
stated to specify a perceptually admissible set rather than the distribution(s)
on it. Interestingly, if no restriction is imposed on how $\hat{S}$
is distributed on the unit circle, then one can simply choose two
antipodal reconstruction points (e.g., those highlighted with $*$
in Fig. \ref{fig:circle}) and perform vector quantization without
involving extra randomness. Indeed, deterministic encoding and decoding
\begin{align*}
 & K:=\begin{cases}
0, & \theta(S)\in[0,\pi),\\
1, & \theta(S)\in[\pi,2\pi),
\end{cases}\\
 & \hat{S}:=\begin{cases}
(0,1), & K=0,\\
(0,-1), & K=1,
\end{cases}
\end{align*}
yields 
\begin{align*}
\mathbb{E}[\|S-\hat{S}\|^{2}]=2-\frac{4}{\pi},
\end{align*}
previously only achievable with the aid of common randomness. Note
that requiring $\hat{S}$ to reside on the unit circle is not a superfluous
``perception constraint" since otherwise it is possible \cite{TA21}
to achieve 
\begin{align*}
\mathbb{E}[\|S-\hat{S}\|^{2}]=1-\frac{4}{\pi^{2}}
\end{align*}
by using the following modified decoder 
\begin{align*}
\hat{S}:=\begin{cases}
(0,\frac{2}{\pi}), & K=0,\\
(0,-\frac{2}{\pi}), & K=1,
\end{cases}
\end{align*}
with the two reconstruction points (highlighted with $\circ$ in Fig.
\ref{fig:circle}) sitting inside the unit circle. Therefore, this
distribution-oblivious ``perception constraint" also requires a
compromisation of distortion. 


By now it should be clear that the role of realization-based perception
constraints is to specify a collection of perceptually admissible
sets parameterized by $P$ (the smaller $P$ is, the more restrictive
the set becomes). One can reconcile such constraints with their distribution-based
counterparts using empirical distribution as a link. To this end,
we quantify the perceptual quality of a given realization $\hat{x}^{n}$
by the divergence $d(p_{X},\gamma_{\hat{x}^{n}})$, where $\gamma_{\hat{x}^{n}}$
is the empirical distribution of $\hat{x}^{n}$. We say $\hat{x}^{n}$
is $P$-typical with respect to $p_{X}$ if $d(p_{X},\gamma_{\hat{x}^{n}})\leq P$,
and let the perceptually admissible set be the set of $P$-typical
sequences. It will be seen that the single-letter characterization
of the rate-distortion-perception function with distribution-based
perception constraints (especially in the form of (\ref{eq:P})) can
be largely recovered in the realization-based framework under the
following definition.

\begin{definition}\label{def:empirical} Given distortion constraint
$D$ and perception constraint $P$, rate $\bar{R}$ is said to be
achievable with common randomness if for all sufficiently large $n$,
there exist shared seed distribution $p_Q$ (on a Polish space), encoding distribution $p_{Z|X^{n}Q}$ (with $\mathcal{Z}$ countable), and
decoding distribution $p_{\hat{X}^{n}|ZQ}$  (with $\hat{\mathcal{X}}=\mathcal{X}$) such that the induced joint distribution
 $p_{X^{n}QZ\hat{X}^{n}}:=p_{X}^{n}p_{Q}p_{Z|X^{n}Q}p_{\hat{X}^{n}|ZQ}$
satisfies 
\begin{align}
 & \frac{1}{n}H(Z|Q)\leq\tilde{R},\nonumber \\
 & \frac{1}{n}\sum\limits _{t=1}^{n}\mathbb{E}[\Delta(X_{t},\hat{X}_{t})]\leq D,\nonumber \\
 & d(p_{X},\gamma_{\hat{X}^{n}})\leq P\mbox{ almost surely}.\label{eq:empiricalperception}
\end{align}
The infimum of such achievable $\bar{R}$ is denoted $\bar{R}_{{\rm {cr}}}(D,P)$.
The achievable rate with private randomness is defined in the same
way except that the encoder and the decoder do not have access to
a shared random seed (i.e., $Q$ is set to be a constant); the corresponding
fundamental limit is denoted $\bar{R}_{{\rm {pr}}}(D,P)$. If the
encoder and the decoder are further required to be deterministic (i.e.,
no randomness is allowed at all), we denote the fundamental limit
by $\bar{R}_{{\rm {nr}}}(D,P)$. \end{definition}

\begin{theorem}\label{eq:empirical} Under Assumptions \ref{assumption:convex} and \ref{assumption:quantization},
\begin{align*}
\bar{R}_{{\rm {cr}}}(D,P)=\bar{R}_{{\rm {pr}}}(D,P)=\bar{R}_{{\rm {nr}}}(D,P)=R(D,P).
\end{align*}
for $D>0$ and $P>0$ \end{theorem} \begin{remark} It is worth mentioning
that $P=0$ is in general not feasible under Definition \ref{def:empirical}
since $p_{X}$ cannot be realized as an empirical distribution for
all sufficiently large $n$. Moreover, $D=0$ implies $\hat{X}^{n}=X^{n}$
almost surely. Note that in the finite alphabet case, $d(p_{X},\gamma_{X^{n}})\leq P$
almost surely for all sufficiently large $n$ if and only if 
\begin{align*}
P\geq\sup\limits _{p_{X'}:p_{X'}\ll p_{X}}d(p_{X},p_{X'}),
\end{align*}
where $p_{X'}\ll p_{X}$ means $p_{\hat{X}'}$ is absolutely continuous
with respect to $p_{X}$. \end{remark} 
\begin{IEEEproof}
First consider the case of common randomness. Let $\bar{R}$ be an
achievable rate with respect to distortion constraint $D$ and perception
constraint $P$. Similarly to (\ref{eq:comb1}) and (\ref{eq:comb2}),
we have 
\begin{align}
 & \bar{R}\geq I(X_{T};\hat{X}_{T}),\label{eq:comb1'}\\
 & D\geq\mathbb{E}[(\Delta(X_{T},\hat{X}_{T}))],\label{eq:comb2'}
\end{align}
where $T$ is uniformly distributed over $[1:n]$ and independent
of $(X^{n},\hat{X}^{n})$. Note that given $\hat{X}^{n}=\hat{x}^{n}$,
the distribution of $\hat{X}_{T}$ is the same as the empirical distribution
of $\hat{x}^{n}$, and consequently 
\begin{align*}
d(p_{X},\gamma_{\hat{x}^{n}})=d(p_{X},p_{\hat{X}_{T}|\hat{X}^{n}=\hat{x}^{n}}).
\end{align*}
Since $d(p_{X},\gamma_{\hat{X}^{n}})\leq P$ almost surely and $d(\cdot,\cdot)$
is convex in its second argument, it follows that 
\begin{align}
d(p_{X},p_{\hat{X}_{T}})\leq P.\label{eq:comb3'}
\end{align}
Combining (\ref{eq:comb1'}), (\ref{eq:comb2'}), (\ref{eq:comb3'})
and invoking the fact that $p_{X_{T}}=p_{X}$ proves $\bar{R}_{{\rm {cr}}}(D,P)\geq R(D,P)$.

Next consider the case of no randomness. According to Assumption \ref{assumption:quantization},
for any $\epsilon\in(0,\min\{\frac{D}{2},\frac{P}{2}\})$, there exists
a discrete random variable $\tilde{X}$ with its support $\tilde{\mathcal{X}}$
satisfying $\tilde{\mathcal{X}}\subseteq\mathcal{X}$ and $|\tilde{\mathcal{X}}|<\infty$
such that 
\begin{align*}
 & I(X;\tilde{X})\leq R(D-2\epsilon,P-2\epsilon)+\epsilon,\\
 & \mathbb{E}[\Delta(X,\tilde{X})]\leq D-\epsilon,\\
 & \mathbb{E}\left[\max\limits _{\tilde{x}\in\tilde{\mathcal{X}}}\Delta(X,\tilde{x})\right]<\infty,\\
 & d(p_{X},p_{\tilde{X}})\leq P-\epsilon.
\end{align*}
By Lemma \ref{lem:deterministic} in Appendix \ref{app:deterministic},
for any $\delta>0$, when $n$ is sufficiently large, we can find
deterministic encoding function $f^{(n)}:\mathcal{X}^{n}\rightarrow[1:M^{(n)}]$
and decoding function $g^{(n)}:[1:M^{(n)}]\rightarrow\mathcal{C}^{(n)}$
with the properties 
\begin{align*}
 & \frac{1}{n}\log M^{(n)}\leq I(X;\tilde{X})+2\delta(H(\tilde{X})+1),\\
 & \frac{1}{n}\sum\limits _{t=1}^{n}\mathbb{E}[\Delta(X_{t},\hat{X}_{t})]\leq\mathbb{E}[\Delta(X,\tilde{X})]+\delta,\\
 & \mathcal{C}^{(n)}\subseteq\mathcal{T}_{\delta}^{(n)}(p_{\tilde{X}}),
\end{align*}
where $\tilde{X}^{n}:=g^{(n)}(f^{(n)}(X^{n}))$ and $\mathcal{T}_{\delta}^{(n)}(p_{\tilde{X}})$
is the set of $\delta$-typical sequences with respect to $p_{\tilde{X}}$.
Clearly, by choosing $\delta\leq\frac{\epsilon}{2(H(\tilde{X})+1)}$,
we have 
\begin{align*}
 & \frac{1}{n}\log M^{(n)}\leq R(D-2\epsilon,P-2\epsilon)+2\epsilon,\\
 & \frac{1}{n}\sum\limits _{t=1}^{n}\mathbb{E}[\Delta(X_{t},\hat{X}_{t})]\leq D.
\end{align*}
Moreover, since $d(p_{X},\gamma)$ is continuous in $\gamma$ over the interior
of the probability simplex defined on $\tilde{\mathcal{X}}$, it follows that
\begin{align*}
d(p_{X},\gamma_{\tilde{x}^{n}})\leq d(p_{X},p_{\tilde{X}})+\epsilon\leq P
\end{align*}
for all $\tilde{x}^{n}\in\mathcal{T}_{\delta}^{(n)}(p_{\tilde{X}})$ when $\delta$
is small enough. Now sending $\epsilon\rightarrow0$ and invoking
the continuity of $R(D,P)$ for $D>0$ and $P>0$  shows $\bar{R}_{{\rm {nr}}}(D,P)\leq R(D,P)$. In view of the
fact that $\bar{R}_{{\rm {cr}}}(D,P)\leq\bar{R}_{{\rm {pr}}}(D,P)\leq\bar{R}_{{\rm {nr}}}(D,P)$,
the proof is complete. 
\end{IEEEproof}
Although defining the perceptual quality of a realization based on
its empirical distribution is arguably justifiable for i.i.d. sources,
this is at best a first-order approximation of what humans subconciously
adopt for image evaluation. How to take into account more sophisticated
patterns while maintaining the theory at a manageable level is a challenge
to be addressed in future works.

\section{Conclusion}

\label{sec:conclusion}

The roles of common randomness and private randomness in rate-distortion-perception
theory have been investigated and shown to depend critically on how
the perception constraint is formulated. The operational meanings
of various perception constraints are also examined. It is our opinion
that, as the distinguishing feature of the new theory and likely the
decisive factor for its success, the notion of perceptual quality
remains to be formalized more convincingly in an information-theoretic
framework as a no-reference metric closely mimicking the relevant
human intuition.

\appendices{}

\section{Verification of Assumption \ref{assumption:quantization}}\label{app:assumptionquantization}

Here we verify Assumption \ref{assumption:quantization} for the case $\mathbb{E}[X^2]<\infty$, $\Delta(x,\hat{x}):=(x-\hat{x})^2$, and $d(p_{X},p_{\hat{X}}):=\inf_{p_{X\hat{X}}\in\Pi(p_{X},p_{\hat{X}})}\mathbb{E}[(X-\hat{X})^{2}]$ (in this case, $d(p_{X},p_{\hat{X}})$ is written as $W^2_2(p_X,p_{\hat{X}})$ according to the convention).

Let $\bar{X}$ be a quantized version of $X$, obtained by mapping
$X$ to its nearest point in $\frac{1}{\sqrt{N}}[-N:N]$, where $N$
is a positive integer. Since $\mathbb{E}[X^2]<\infty$, for any $D>0$, we can choose a sufficiently large $N$ such that $\mathbb{E}[(X-\bar{X})^2]\leq\frac{D}{2}$.
Let $\hat{X}$ be the mirror version of $X$ with respect to $\check{X}:=\mathbb{E}[X|\bar{X}]$ in the sense that $p_{X\check{X}\hat{X}}=p_{X\check{X}}p_{\hat{X}|\check{X}}$ and $p_{X|\check{X}}=p_{\hat{X}|\check{X}}$.
 Obviously, $X\leftrightarrow \check{X}\leftrightarrow\hat{X}$
form a Markov chain and $p_{\hat{X}}=p_{X}$. Note that
\begin{align*}
&I(X;\hat{X})\leq I(X;\check{X})\leq H(\check{X})\leq H(\bar{X})\leq\log(2N+1),\\
&\mathbb{E}[(X-\hat{X})^2]=2\mathbb{E}[(X-\check{X})^2]\leq 2\mathbb{E}[(X-\bar{X})^2]\leq D,\\
&W^2_2(p_X,p_{\hat{X}})=0.
\end{align*}
Therefore, we have\footnote{It is worth mentioning that the value of $R(D,0)$ does not depend on the choice of divergence $d$.} $R(D,0)<\infty$. Since $R(D,P)\leq R(D,0)$ for $P>0$, it follows that  (\ref{eq:finiteR}) holds.

By the definition of $R(D,P)$, for any $D>0$ and $P>0$, there exists a random
variable $\hat{X}$ such that 
\begin{align*}
& I(X;\hat{X})\leq R(D,P)+\epsilon,\\
& \mathbb{E}[(X-\hat{X})^2]\leq D,\\
& W^2_2(p_{X},p_{\hat{X}})\leq P.
\end{align*}
Let $\tilde{X}$ be a quantized version of $\hat{X}$, obtained by mapping
$X$ to its nearest point in $\frac{1}{\sqrt{N}}[-N:N]$, where $N$
is a positive integer. Clearly, we have $|\tilde{\mathcal{X}}|<\infty$ since $\tilde{\mathcal{X}}\subseteq \frac{1}{\sqrt{N}}[-N:N]$. With this construction, (\ref{eq:app1}) is a simple consequence of the data processing inequality. It can be verified that
\begin{align}
&\mathbb{E}[(X-\tilde{X})^2]\\
&=\mathbb{E}[(X-\hat{X}+\hat{X}-\tilde{X})^2]\nonumber\\
&=\mathbb{E}[(X-\hat{X})^2]+\mathbb{E}[(\hat{X}-\tilde{X})^2]\nonumber\\
&\quad+2\mathbb{E}[(X-\hat{X})(\hat{X}-\tilde{X})]\nonumber\\
&\leq\mathbb{E}[(X-\hat{X})^2]+\mathbb{E}[(\hat{X}-\tilde{X})^2]\nonumber\\
&\quad+2\sqrt{\mathbb{E}[(X-\hat{X})^2]\mathbb{E}[(\hat{X}-\tilde{X})^2]}\label{eq:CS}\\
&\leq D+\mathbb{E}[(\hat{X}-\tilde{X})^2]+2\sqrt{D\mathbb{E}[(\hat{X}-\tilde{X})^2]},\nonumber
\end{align}
where (\ref{eq:CS}) is due to the Cauchy-Schwarz inequality.
Similarly,
\begin{align*}
W^2_2(p_X,p_{\tilde{X}})&\leq P+\mathbb{E}[(\hat{X}-\tilde{X})^2]+2\sqrt{P\mathbb{E}[(\hat{X}-\tilde{X})^2]}.
\end{align*}
Since $\mathbb{E}[\hat{X}^2]<\infty$ (implied by the fact $\mathbb{E}[X^2]<\infty$ and $\mathbb{E}[(X-\hat{X})^2]<\infty$), it follows that $\mathbb{E}[(\hat{X}-\tilde{X})^2]\rightarrow 0$ as $N\rightarrow\infty$. Therefore, we can ensure (\ref{eq:app2}) and (\ref{eq:app4}) by setting $N$ large enough. Moreover, (\ref{eq:app3}) is satisifed because
\begin{align}
\mathbb{E}\left[\max\limits_{\tilde{x}\in\tilde{\mathcal{X}}}(X-\tilde{x})^2\right]&\leq\mathbb{E}[X^2]+2\sqrt{N}\mathbb{E}[|X|]+N\nonumber\\
&\leq\mathbb{E}[X^2]+2\sqrt{N\mathbb{E}[X^2]}+N\label{eq:CS2}\\
&<\infty,\nonumber
\end{align}
where (\ref{eq:CS2}) is due to the Cauchy-Schwarz inequality.
In view of the fact that 
$W^2_2(p_X,\gamma)\leq\mathbb{E}[\max_{\tilde{x}\in\tilde{\mathcal{X}}}(X-\tilde{x})^2]$ for all $\gamma$ with support $\tilde{\mathcal{X}}$, (\ref{eq:finited}) must hold as well. This completes the verification of Assumption \ref{assumption:quantization}.

\section{Proof of Theorem \ref{thm:deterministic}}

\label{app:deterministic}

\begin{lemma}\label{lem:deterministic} Let $\tilde{X}$ be a discrete random
variable defined on  $\mathcal{X}$ and jointly distributed with $X$ and assume that its support
$\tilde{\mathcal{X}}\subseteq\mathcal{X}$ satisfies 
$|\tilde{\mathcal{X}}|<\infty$ and 
\begin{align*}
\mathbb{E}\left[\max\limits _{\tilde{x}\in\tilde{\mathcal{X}}}\Delta(X,\tilde{x})\right]<\infty.
\end{align*}
Given any $D>0$, $P>0$, and $\delta>0$, there exist deterministic
encoding function $f^{(n)}:\mathcal{X}^{n}\rightarrow[1:M^{(n)}]$
and decoding function $g^{(n)}:[1:M^{(n)}]\rightarrow\mathcal{C}^{(n)}$
for all sufficiently large $n$ such that 
\begin{align*}
 & \frac{1}{n}\log M^{(n)}\leq I(X;\tilde{X})+2\delta(H(\tilde{X})+1),\\
 & \frac{1}{n}\sum\limits _{t=1}^{n}\mathbb{E}[\Delta(X_{t},\tilde{X}_{t})]\leq\mathbb{E}[\Delta(X,\tilde{X})]+\delta,\\
 & \mathcal{C}^{(n)}\subseteq\mathcal{T}_{\delta}^{(n)}(p_{\tilde{X}}),
\end{align*}
where $\tilde{X}^{n}:=g^{(n)}(f^{(n)}(X^{n}))$ and $\mathcal{T}_{\delta}^{(n)}(p_{\tilde{X}})$
denotes the set of $\delta$-typical sequences with respect to $p_{\tilde{X}}$,
i.e., $\mathcal{T}_{\delta}^{(n)}(p_{\tilde{X}}):=\{\tilde{x}^{n}\in\tilde{\mathcal{X}}^{n}:|\gamma_{\tilde{x}^{n}}(\tilde{x})-p_{\tilde{X}}(\tilde{x})|\leq\delta p_{\tilde{X}}(\tilde{x})\mbox{ for all }\tilde{x}\in\tilde{\mathcal{X}}\}$.
\end{lemma} 
\begin{IEEEproof}
This result has many known variants in the literature (see, e.g.,
\cite[Theorem 9.6.2]{Gallager68}). We include its proof for completeness.

We indepenently and uniformly choose $M^{(n)}:=\lfloor2^{n(I(X;\tilde{X})+2\delta')}\rfloor$
codewords $\mathcal{C}^{(n)}:=\{\tilde{X}^{n}\}_{m=1}^{M^{(n)}}$
from $\mathcal{T}_{\delta}^{(n)}(p_{\tilde{X}})$. Given $X^{n}=x^{n}$, the encoder
finds an $\hat{m}$ such that 
\begin{align}
\frac{1}{n}\sum\limits _{t=1}^{n}\Delta(x_{t},\tilde{X}_{t}(\hat{m}))\leq\mathbb{E}[\Delta(X,\tilde{X})]+\frac{\delta}{2}.\label{eq:distortion}
\end{align}
If no such $\hat{m}$ exists, the encoder simply sets $\hat{m}=1$.
Let 
\begin{align*}
\mathcal{A}_{x^{n}}:= & \left\{ \tilde{x}^{n}\in\mathcal{T}_{\delta}^{(n)}(p_{\tilde{X}}):\prod\limits _{t=1}^{n}p_{\tilde{X}|X}(\tilde{x}_{t}|x_{t})>\frac{2^{n\hat{R}}}{|\mathcal{T}_{\delta}^{(n)}(p_{\tilde{X}})|}\right.\\
 & \left.\mbox{ or }\frac{1}{n}\sum\limits _{t=1}^{n}\Delta(x_{t},\tilde{x}_{t})>\mathbb{E}[\Delta(X,\tilde{X})]+\frac{\delta}{2}\right\} ,
\end{align*}
where $\hat{R}:=I(X;\tilde{X})+\delta'$. Given $X^{n}=x^{n}$, the
encoder fails to find a codeword satisfying (\ref{eq:distortion})
only if $\tilde{X}^{n}(m)\in\mathcal{A}_{x^{n}}$ for all $m$. Therefore,
\begin{align}
 & \mathbb{P}\Bigg(\frac{1}{n}\sum\limits _{t=1}^{n}\Delta(x_{t},\tilde{X}^{n}(\hat{m}))\nonumber \\
 & \quad\left.>\mathbb{E}[\Delta(X,\tilde{X})]+\frac{\delta}{2}\mbox{ for all }m\right|X^{n}=x^{n}\Bigg)\nonumber \\
 & \leq\left(1-\sum_{\tilde{x}^{n}\in\mathcal{T}_{\delta}^{(n)}(p_{\tilde{X}})\backslash\mathcal{A}_{x^{n}}}\frac{1}{|\mathcal{T}_{\delta}^{(n)}(p_{\tilde{X}})|}\right)^{M^{(n)}}\nonumber \\
 & \leq\left(1-2^{-n\hat{R}}\sum_{\tilde{x}^{n}\in\mathcal{T}_{\delta}^{(n)}(p_{\tilde{X}})\backslash\mathcal{A}_{x^{n}}}\prod\limits _{t=1}^{n}p_{\tilde{X}|X}(\tilde{x}_{t}|x_{t})\right)^{M^{(n)}}\nonumber \\
 & \leq1-\sum_{\tilde{x}^{n}\in\mathcal{T}_{\delta}^{(n)}(p_{\tilde{X}})\backslash\mathcal{A}_{x^{n}}}\prod\limits _{t=1}^{n}p_{\tilde{X}|X}(\tilde{x}_{t}|x_{t})\nonumber\\
 &\quad+\exp(-M^{(n)}2^{-n\hat{R}})\label{eq:inequality} \\
 & =\sum_{\tilde{x}^{n}\in(\mathcal{T}_{\delta}^{(n)}(p_{\tilde{X}}))^{c}\cup\mathcal{A}_{x^{n}}}\prod\limits _{t=1}^{n}p_{\tilde{X}|X}(\tilde{x}_{t}|x_{t})\nonumber\\
 &\quad+\exp(-M^{(n)}2^{-n\hat{R}}),\label{eq:conditional}
\end{align}
where (\ref{eq:inequality}) is due to \cite[Lemma 13.5.3]{CT91}.
Let $\mathcal{A}^{(n)}:=\{(x^{n},\tilde{x}^{n}):x^{n}\in\mathcal{X}^{n},\tilde{x}^{n}\in(\mathcal{T}_{\delta}^{(n)}(p_{\tilde{X}}))^{c}\cup\mathcal{A}_{x^{n}}\}$.
It follows from (\ref{eq:conditional}) that 
\begin{align*}
 & \mathbb{P}\left(\frac{1}{n}\sum\limits _{t=1}^{n}\Delta(X_{t},\tilde{X}(\hat{m}))>\mathbb{E}[\Delta(X,\tilde{X})]+\frac{\delta}{2}\right)\\
 & \leq\mathbb{P}((X^{n},\check{X}^{n})\in\mathcal{A}^{(n)})+\exp(-M^{(n)}2^{-n\hat{R}}),
\end{align*}
where $(X_{t},\check{X}_{t})$, $t\in[1:n]$, are independent and
distributed according to $p_{X,\tilde{X}}$. Note that $\check{X}^{n}\in\mathcal{T}_{\delta}^{(n)}(p_{\tilde{X}})$
with high probability when $n$ is large. Moreover, in light of the
weak law of large numbers, $\frac{1}{n}\sum_{t=1}^{n}p_{\tilde{X}|X}(\check{X}_{t}|X_{t})$
and $\frac{1}{n}\sum_{t=1}^{n}\Delta(X_{t},\check{X}_{t})$ converge,
respectively, to $-H(\tilde{X}|X)$ and $\mathbb{E}[\Delta(X,\tilde{X})]$
in probability. It is easy to see that 
\begin{align*}
\frac{1}{n}\log\frac{2^{n\hat{R}}}{|\mathcal{T}_{\delta}^{(n)}(p_{\tilde{X}})|} & \geq\frac{1}{n}\log\frac{2^{n\hat{R}}}{2^{n(1+\delta)H(\tilde{X})}}\\
 & =I(X;\tilde{X})+\delta'-(1+\delta)H(\tilde{X})\\
 & =-H(\tilde{X}|X)+\delta'-\delta H(\tilde{X}).
\end{align*}
So by choosing $\delta'=\delta(H(\tilde{X})+1)$, we have $\mathbb{P}((X^{n},\check{X}^{n})\in\mathcal{A}^{(n)})\rightarrow0$
as $n\rightarrow\infty$. It is also clear that $\exp(-M^{(n)}2^{-n\hat{R}})\rightarrow0$
as $n\rightarrow\infty$.

As shown by the above argument, for any $\delta>0$ and $\epsilon>0$,
when $n$ is sufficiently large, we can find deterministic encoding
function $f^{(n)}:\mathcal{X}^{n}\rightarrow[1:M^{(n)}]$ and decoding
function $g^{(n)}:[1:M^{(n)}]\rightarrow\mathcal{C}^{(n)}$ such that
\begin{align*}
 & \frac{1}{n}\log M^{(n)}\leq I(X;\tilde{X})+2\delta(H(\tilde{X})+1),\\
 & \mathbb{P}\left(\frac{1}{n}\sum\limits _{t=1}^{n}\Delta(X_{t},\hat{X}_{t})>\mathbb{E}[\Delta(X,\tilde{X})]+\frac{\delta}{2}\right)\leq\epsilon,\\
 & \mathcal{C}^{(n)}\subseteq\mathcal{T}_{\delta}^{(n)}(p_{\tilde{X}}).
\end{align*}
Let 
\begin{align*}
W:=\begin{cases}
1, & \frac{1}{n}\sum\limits _{t=1}^{n}\Delta(X_{t},\hat{X}_{t})>\mathbb{E}[\Delta(X,\tilde{X})]+\frac{\delta}{2},\\
0, & \mbox{otherwise},
\end{cases}
\end{align*}
and $V_{t}:=\max_{\tilde{x}\in\tilde{\mathcal{X}}}\Delta(X_{t},\tilde{x})$,
$t\in[1:n]$. We have 
\begin{align*}
 & \frac{1}{n}\sum\limits _{t=1}^{n}\mathbb{E}[\Delta(X_{t},\hat{X}_{t})]\\
 & =\frac{1}{n}\sum\limits _{t=1}^{n}\mathbb{P}(W=0)\mathbb{E}[\Delta(X_{t},\hat{X}_{t})|W=0]\\
 & \quad+\frac{1}{n}\sum\limits _{t=1}^{n}\mathbb{P}(W=1)\mathbb{E}[\Delta(X_{t},\hat{X}_{t})|W=1]\\
 & \leq\frac{1}{n}\sum\limits _{t=1}^{n}\mathbb{E}[\Delta(X_{t},\hat{X}_{t})|W=0]\\
 & \quad+\frac{1}{n}\sum\limits _{t=1}^{n}\mathbb{E}[W\Delta(X_{t},\hat{X}_{t})]\\
 & \leq\mathbb{E}[\Delta(X,\tilde{X})]+\frac{\delta}{2}+\frac{1}{n}\sum\limits _{t=1}^{n}\mathbb{E}[WV_{t}].
\end{align*}
Note that 
\begin{align*}
\mathbb{E}[WV_{t}] & =\mathbb{P}\left(V_{t}\leq\frac{1}{\sqrt{\epsilon}}\right)\mathbb{E}\left[WV_{t}\left|V_{t}\leq\frac{1}{\sqrt{\epsilon}}\right.\right]\\
 & \quad+\mathbb{P}\left(V_{t}>\frac{1}{\sqrt{\epsilon}}\right)\mathbb{E}\left[WV_{t}\left|V_{t}>\frac{1}{\sqrt{\epsilon}}\right.\right]\\
 & \leq\mathbb{P}\left(V_{t}\leq\frac{1}{\sqrt{\epsilon}}\right)\mathbb{E}\left[W\frac{1}{\sqrt{\epsilon}}\left|V_{t}\leq\frac{1}{\sqrt{\epsilon}}\right.\right]\\
 & \quad+\mathbb{P}\left(V_{t}>\frac{1}{\sqrt{\epsilon}}\right)\mathbb{E}\left[V_{t}\left|V_{t}>\frac{1}{\sqrt{\epsilon}}\right.\right]\\
 & \leq\frac{1}{\sqrt{\epsilon}}\mathbb{E}[W]+\mathbb{P}\left(V_{t}>\frac{1}{\sqrt{\epsilon}}\right)\mathbb{E}\left[V_{t}\left|V_{t}>\frac{1}{\sqrt{\epsilon}}\right.\right]\\
 & \leq\sqrt{\epsilon}+\mathbb{P}\left(V_{t}>\frac{1}{\sqrt{\epsilon}}\right)\mathbb{E}\left[V_{t}\left|V_{t}>\frac{1}{\sqrt{\epsilon}}\right.\right].
\end{align*}
Since 
\begin{align*}
\mathbb{E}[V_{t}]=\mathbb{E}\left[\max\limits _{\tilde{x}\in\tilde{\mathcal{X}}}\Delta(X,\tilde{x})\right]<\infty,
\end{align*}
it follows by the dominated convergence theorem that $\mathbb{P}(V_{t}>v)\mathbb{E}[V_{t}|V_{t}>v]\rightarrow0$
as $v\rightarrow\infty$. Therefore, by choosing a sufficiently small
$\epsilon$, we can ensure that $\mathbb{E}[WV_{t}]\leq\frac{\delta}{2}$,
$t\in[1:n]$, and consequently 
\begin{align*}
\frac{1}{n}\sum\limits _{t=1}^{n}\mathbb{E}[\Delta(X_{t},\hat{X}_{t})]\leq\mathbb{E}[\Delta(X,\tilde{X})]+\delta.
\end{align*}
This completes the proof of Lemma \ref{lem:deterministic}. 
\end{IEEEproof}
Now we proceed to prove Theorem \ref{thm:deterministic}. It suffices
to consider the case where $p_{X}$ does not assign all probability
mass to a single atom since otherwise the problem is trivial.

According to Assumption \ref{assumption:quantization}, for any $\epsilon\in(0,\min\{\frac{D}{2},\frac{P}{2}\})$,
there exists a random variable $\tilde{X}$ with its support $\tilde{\mathcal{X}}$
satisfying $\tilde{\mathcal{X}}\subseteq\mathcal{X}$ and $|\tilde{\mathcal{X}}|<\infty$
such that 
\begin{align}
 & I(X;\tilde{X})\leq R(D-2\epsilon,P-2\epsilon)+\epsilon,\nonumber \\
 & \mathbb{E}[\Delta(X,\tilde{X})]\leq D-\epsilon,\nonumber \\
 & \mathbb{E}\left[\max\limits _{\tilde{x}\in\tilde{\mathcal{X}}}\Delta(X,\tilde{x})\right]<\infty,\label{eq:maxdistortion}\\
 & d(p_{X},p_{\tilde{X}})\leq P-\epsilon.\nonumber 
\end{align}
By Lemma \ref{lem:deterministic}, for any $\delta>0$, when $n$
is sufficiently large, we can find deterministic encoding function
$f^{(n)}:\mathcal{X}^{n}\rightarrow[1:M^{(n)}]$ and decoding function
$g^{(n)}:[1:M^{(n)}]\rightarrow\mathcal{C}^{(n)}$ with the properties
\begin{align*}
 & \frac{1}{n}\log M^{(n)}\leq I(X;\tilde{X})+2\delta(H(\tilde{X})+1),\\
 & \frac{1}{n}\sum\limits _{t=1}^{n}\mathbb{E}[\Delta(X_{t},\tilde{X}_{t})]\leq\mathbb{E}[\Delta(X,\tilde{X})]+\delta,\\
 & \mathcal{C}^{(n)}\subseteq\mathcal{T}_{\delta}^{(n)}(p_{\tilde{X}}).
\end{align*}
Clearly, by choosing $\delta\leq\frac{\epsilon}{2(H(\tilde{X})+1)}$,
we have 
\begin{align*}
 & \frac{1}{n}\log M^{(n)}\leq R(D-2\epsilon,P-2\epsilon)+2\epsilon,\\
 & \frac{1}{n}\sum\limits _{t=1}^{n}\mathbb{E}[\Delta(X_{t},\tilde{X}_{t})]\leq D-\frac{\epsilon}{2}.
\end{align*}
Moreover, $\frac{1}{n}\sum_{t=1}^{n}p_{\tilde{X}_{t}}$ must converge
to $p_{\tilde{X}}$ under the total variation  distance (i.e., $d_{\mathrm{TV}}(\frac{1}{n}\sum_{t=1}^{n}p_{\tilde{X}_{t}},p_{\tilde{X}})\rightarrow 0$)
as $\delta\rightarrow0$
given the fact that 
\begin{align*}
 & \frac{1}{n}\sum\limits _{t=1}^{n}p_{\tilde{X}_{t}|\tilde{X}^{n}=\tilde{x}^{n}}=\gamma_{\tilde{x}^{n}},\\
 & \mathbb{P}(\tilde{X}^{n}\in\mathcal{T}_{\delta}^{(n)}(p_{\tilde{X}}))=1.
\end{align*}
Since $d(p_{X},\gamma)$ is continuous in $\gamma$ over the interior of the
probability simplex defined on $\tilde{\mathcal{X}}$, it follows that
\begin{align}
d\left(p_{X},\frac{1}{n}\sum\limits _{t=1}^{n}p_{\tilde{X}_{t}}\right)\leq d(p_{X},p_{\tilde{X}})+\frac{\epsilon}{2}\leq P-\frac{\epsilon}{2}\label{eq:div_ave}
\end{align}
when $\delta$ is sufficiently close to zero.

However, in general (\ref{eq:div_ave}) does not imply the stronger
requirement 
\begin{align*}
d(p_{X},p_{\tilde{X}_{t}})\leq P-\frac{\epsilon}{2},\quad t\in[1:n].
\end{align*}
Nevertheless, there is a simple remedy with the availability of common
randomness. For any integer $q$, let $s_{i}^{(n)}$ be a circular
shift operator in the sense that $s_{q}^{(n)}(x^{n})=(x_{1\oplus_{n}q},x_{2\oplus_{n}q},\cdots,x_{n\oplus_{n}q})$
for all $x^{n}$, where $\oplus_{n}$ is modulo-$n$ addition\footnote{We assume the output of the modulo-$n$ operation is in $[1:n]$.}.
Let $\mathcal{C}_{q}^{(n)}$ denote the codebook obtained by applying
$s_{q}^{(n)}$ to every codeword\footnote{The resulting codeword still retains the index of the original codeword.}
of $\mathcal{C}^{(n)}$, $q\in[0:n-1]$. Moreover, we equip each codebook
$\mathcal{C}_{q}^{(n)}$ with encoding function $f_{q}^{(n)}:\mathcal{X}^{n}\rightarrow[1:M^{(n)}]$
and decoding function $g_{q}^{(n)}:[1:M^{(n)}]\rightarrow\mathcal{C}_{q}^{(n)}$
defined as follows: 
\begin{align*}
 & f_{q}^{(n)}(x^{n}):=f^{(n)}(s_{-q}^{(n)}(x^{n})),\quad x^{n}\in\mathcal{X}^{n},\\
 & g_{q}^{(n)}(m)=s_{q}^{(n)}(g^{(n)}(m)),\quad m\in[1:M^{(n)}].
\end{align*}
Note that we have $\mathcal{C}_{0}^{(n)}=\mathcal{C}^{(n)}$, $f_{0}^{(n)}=f^{(n)}$,
and $g_{0}^{(n)}=g^{(n)}$. Let $Q$ be uniformly distributed over
$[0:n-1]$, independent of $X^{n}$, and available at both the encoder
and decoder. The role of $Q$ is to specify which codebook (as well
as the associated encoding and decoding functions) to use. Let $\check{X}^{n}$
denote the reconstruction. Based on our construction, it is clear
that 
\begin{align*}
 & p_{\check{X}_{t}}=\frac{1}{n}\sum\limits _{t'=1}^{n}p_{\tilde{X}_{t'}},\quad t\in[1:n],\\
 & \frac{1}{n}\sum\limits _{t=1}^{n}\mathbb{E}[\Delta(X_{t},\check{X}_{t})|Q=q]=\frac{1}{n}\sum\limits _{t=1}^{n}\mathbb{E}[\Delta(X_{t},\tilde{X}_{t})],\\
 & \hspace{2in}q\in[0:n-1],
\end{align*}
and consequently 
\begin{align*}
 & d(p_{X},p_{\check{X}_{t}})\leq P-\frac{\epsilon}{2},\quad t\in[1:n],\\
 & \frac{1}{n}\sum\limits _{t=1}^{n}\mathbb{E}[\Delta(X_{t},\check{X}_{t})]\leq D-\frac{\epsilon}{2}.
\end{align*}

It remains to eliminate common randomness. The key observation here
is that $Q$ can be simulated using a negligible fraction of source
symbols as $H(Q)$ is sublinear in $n$. Specificaly, consider the
case where $n_{0}:=\lfloor n\alpha\rfloor$ source symbols are leveraged
to simulate $Q$, where $\alpha>0$ is small enough. Let $p_{\max}$
denote the maximum probability value of $p_{X}$ (i.e., the maximum
of the probabilities of atoms if atoms exist; otherwise, $p_{\max}=0$).
We assume $p_{\max}>0$ since otherwise $Q$ can be exactly generated
from $X\sim p_{X}$. Obviously, the maximum probability value of $p_{X}^{n_{0}}$
is $p_{\max}^{n_{0}}$, which vanishes exponentially as $n\to\infty$
(for any given $\alpha>0$). This ensures that there is a map $\omega:\mathcal{X}^{n_{0}}\to[0:n-1]$
such that the distribution $p_{\omega(X^{n_{0}})}$ of $\omega(X^{n_{0}})$
satisfies $|p_{\omega(X^{n_{0}})}(i)-\frac{1}{n}|\le p_{\max}^{n_{0}}$
for all $i\in[0:n-1]$. 
For example, first map as many atoms as possible to the elements of
$[0:n-1]$ while ensuring that the total mass of atoms mapped to each
element is no greater than $\frac{1}{n}$; then use the remaining
atoms and atomless measurable sets to fill the gap between the total
mass for each element and the target $\frac{1}{n}$ so that the total
mass for each element is within the range $\frac{1}{n}\pm p_{\max}^{n_{0}}$.
Consequently, we have $d_{\mathrm{TV}}(p_{\omega(X^{n_{0}})},\mathrm{Unif}[0:n-1])\le np_{\max}^{n_{0}}\to0$
as $n\to\infty$.

We assume the encoder uses this simulation code on the $(n+1)$-th
to $(n+n_{0})$-th source symbols $X_{n+1}^{n+n_{0}}$ to generate
$\hat{Q}:=\omega(X_{n+1}^{n+n_{0}})$. It then transmits $\hat{Q}$
to the decoder losslessly with $\left\lceil \log n\right\rceil $
bits and applies the aforementioned randomly shifted code to the first
$n$ source symbols $X^{n}$ but with $Q$ replaced by its approximate
version $\hat{Q}$. Denote the reconstructions of the first $n$ source
symbols by $\hat{X}^{n}$. For the $(n+1)$-th to $(n+n_{0})$-th
source symbols, the decoder generates reconstructions $\hat{X}_{n+j}:=\hat{X}_{j}$,
$j\in[1:n_{0}]$. Under this construction, both the encoder and decoder
are deterministic.

We shall show that the deterministic code has the desired properties.
Since 
\begin{align*}
 & \sum\limits _{t=1}^{n}\mathbb{E}[\Delta(X_{t},\hat{X}_{t})|Q=q]=\sum\limits _{t=1}^{n}\mathbb{E}[\Delta(X_{t},\tilde{X}_{t})],\\
 & \hspace{2in}q\in[0:n-1],\\
 & \mathbb{E}[\Delta(X_{n+j},\hat{X}_{n+j})]\leq\mathbb{E}\left[\max\limits _{\tilde{x}\in\tilde{\mathcal{X}}}\Delta(X,\tilde{x})\right]<\infty,\\
 & \hspace{2in}j\in[1:n_{0}],
\end{align*}
it follows that 
\begin{align*}
\frac{1}{n+n_{0}}\sum\limits _{t=1}^{n+n_{0}}\mathbb{E}[\Delta(X_{t},\hat{X}_{t})]\leq D
\end{align*}
for all sufficiently large $n$ as long as $\alpha$ is chosen small
enough. Moreover, in view of the fact that $p_{\hat{X}_{t}|\hat{Q}}=p_{\check{X}_{t}|Q}$,
$t\in[1:n]$, and $d_{\mathrm{TV}}(p_{\hat{Q}},p_{Q})\rightarrow0$ as $n\rightarrow\infty$,
we must have $d_{\mathrm{TV}}(p_{\hat{X}_{t}},p_{\check{X}_{t}})\rightarrow0$
uniformly for all $t\in[1:n]$ as $n\rightarrow\infty$, which, together
with the uniform convergence of $p_{\check{X}_{t}}$, $t\in[1:n]$,
to $p_{\tilde{X}}$ under the total variation distance as $\delta\rightarrow0$
and the continuity of $d(p_{X},q)$ in this second argument at $q=p_{\tilde{X}}$,
implies 
\begin{align*}
d(p_{X},p_{\check{X}_{t}})\leq P,\quad t\in[1:n+n_{0}],
\end{align*}
for all sufficiently large $n$ as long as $\delta$ is chosen small
enough. Finally, invoking the fact that $\frac{\lceil\log n\rceil}{n+n_{0}}\rightarrow0$
as $n\rightarrow\infty$ and the continuity of $R(D,P)$ for $D>0$
and $P>0$  completes the proof of Theorem
\ref{thm:deterministic}.

\begin{remark} It is possible to avoid using the simulation code
via a more careful construction.

Let $\mathcal{A}:=\{x^{n}\in\mathcal{X}^{n}:x^{n}=s_{i}^{(n)}(x^{n})\mbox{ for some }i\in[1:n-1]\}$
and $\mathcal{B}:=\mathcal{X}^{n}\backslash\mathcal{A}$. Moreover,
we partition $\mathcal{B}$ into $\mathcal{B}_{0},\mathcal{B}_{1},\cdots,\mathcal{B}_{n-1}$
such that $\mathcal{B}_{q}=\{s_{i}^{(n)}(x^{n}):x^{n}\in\mathcal{B}_{0}\}$,
$q\in[1:n-1]$. (The existence of such a partition will be discussed later.)

Take codebook $\mathcal{C}^{n}\subseteq\mathcal{T}_{\delta}^{(n)}(p_{\tilde{X}})$
and its associated deterministic encoding function $f^{(n)}:\mathcal{X}^{n}\rightarrow[1:M^{(n)}]$
and decoding function $g^{(n)}:[1:M^{(n)}]\rightarrow\mathcal{C}^{(n)}$
as specified in the proof of Theorem \ref{thm:deterministic}. Given
$x^{n}\in\mathcal{A}\cup\mathcal{B}_{0}$, the encoder finds\footnote{Use a prescribed deterministic tie-break rule if the minimizer is
not unique.} $(m^{*},q^{*})$ such that 
\begin{align*}
(m^{*},q^{*})=\arg\min\limits _{m\in[1:M^{(n)}],i\in[0:n-1]}\sum\limits _{t=1}^{n}\Delta(x_{t},\tilde{x}_{t}),
\end{align*}
where $\tilde{x}^{n}:=s_{q}^{(n)}(g^{(n)}(m))$. It then sends $(m^{*},q^{*})$
to the decoder, which produces $s_{i^{*}}^{(n)}(g^{(n)}(m^{*}))$
as the reconstruction. For $x^{n}\in\mathcal{B}_{q}$, $q\in[1:n-1]$,
the encoder finds the index pair for the corresponding $s_{-q}^{(n)}(x^{n})$
in $\mathcal{B}_{0}$ according to the aforedescribed encoding rule
and sends it to the decoder. The decoder first produces the reconstruction
for $s_{-q}^{(n)}(x^{n})$ according to the aforedescribed decoding
rule, then outputs the final result by shifting this reconstruction
using $s_{q}^{(n)}$. Let $\bar{X}^{n}$ denote the output induced
by this new coding scheme. Since the distortion does not increase
for any realization $x^{n}$ as compared to the original scheme\footnote{If we modify the coding scheme by setting $(m^{*},q^{*})=(f^{(n)}(x^{n}),0)$
for $x^{n}\in\mathcal{B}_{0}$, then the distortion might increase
for $x^{n}\in\mathcal{B}\backslash\mathcal{B}_{0}$.}, it follows that 
\begin{align}
\sum\limits _{t=1}^{n}\mathbb{E}[\Delta(X_{t},\bar{X}_{t})]\leq\sum\limits _{t=1}^{n}\mathbb{E}[\Delta(X_{t},\tilde{X}_{t})],\label{eq:dominance}
\end{align}
where $\tilde{X}^{n}:=g^{(n)}(f^{(n)}(X^{n}))$. Moreover, as shown
at the end of Appendix \ref{app:deterministic}, 
\begin{align}
\lim\limits _{n\rightarrow\infty}\mathbb{P}(X^{n}\in\mathcal{A})=0.\label{eq:shift}
\end{align}
So with $\delta$ chosen small enough, the circular symmetry of the
coding scheme (conditioned on $X^{n}\in\mathcal{B}$) ensures that
$p_{\bar{X}_{t}}$, $t\in[1:n]$, are uniformly close to $p_{\tilde{X}}$
under the total variation distance and consequently 
\begin{align*}
d(p_{X},p_{\bar{X}_{t}})\leq P,\quad t\in[1:n],
\end{align*}
for all sufficiently large $n$. As the rate overhead for transmitting
$q^{*}$ is negligible, the new coding scheme indeed has the desired
properties.

However, there is a subtle issue regarding the partition of $\mathcal{B}$
into $\mathcal{B}_{0},\mathcal{B}_{1},\cdots,\mathcal{B}_{n-1}$.
The existence of such a partition is in a certain sense guaranteed.
Note that $\mathcal{B}$ is the union of a collection of disjoint
equivalent classes, where each equivalent class consists of $n$ different
sequences that can be converted from one to another via circular shifting.
We can form $\mathcal{B}_{0}$ by choosing one sequence from each
equivalent class in $\mathcal{B}$; then $\mathcal{B}_{i}$ is uniquely
specified due to the requirement $\mathcal{B}_{q}=\{s_{q}^{(n)}(x^{n}):x^{n}\in\mathcal{B}_{0}\}$,
$q\in[1:n-1]$. It can be seen that there is considerable freedom
in creating this kind of partitions. When $\mathcal{X}$ is finite
or countably infinite, the resulting sets are always measurable. But
in a more general setting, certain partitions might yield non-measurable
sets.

Here we show that this issue can be resolved by performing the partition judiciously when $\mathcal{X}$ is a Polish space (as assumed throughout this paper). 
First consider the case $\mathcal{X}=\mathbb{R}$.
We shall start with a simple example
where $n=3$. There are totally $6$ permutations on $[1:3]$, namely,
$(1,2,3)$, $(1,3,2)$, $(2,1,3)$, $(2,3,1)$, $(3,1,2)$, and $(3,2,1)$.
Pick two permutations that are not related via circular shifting\footnote{For example, $(1,2,3)$ and $(1,3,2)$ are not related via circular
shifting while $(1,2,3)$ and $(3,1,2)$ are since applying $s_{2}^{(3)}$
to $(1,2,3)$ gives $(3,1,2)$.}. Each permutation $(a,b,c)$ can be used to specificy a region in
$\mathbb{R}^{3}$ according to the following rule: $(a,b,c)\mapsto \{x^{3}\in\mathbb{R}^{3}:x_{a}\geq x_{b}\geq x_{c}\}$.
Now we can let $\mathcal{B}_{0}$ be the union of the regions specified
by the two picked permutations 
with the elements in $\mathcal{A}$
excluded. Then $\mathcal{B}_{1}$ and $\mathcal{B}_{2}$ are also
uniquely determined. In general, there are $n!$ permutations on $[1:n]$.
We can pick $(n-1)!$ permutations that are not related via circular
shifting, and use these permutations to specify $(n-1)!$ regions
in $\mathbb{R}^{n}$ (according to the obvious extension of the aforementioned
rule). Then define $\mathcal{B}_{0}$ by taking the union of these
$(n-1)!$ regions and excluding the elements in $\mathcal{A}$. It
can be verified that the induced partition $\mathcal{B}_{0},\mathcal{B}_{1},\cdots,\mathcal{B}_{n-1}$
is free of the measurability issue.

This method can be generalized  to  any  Polish space. Note that  any Borel space on a Polish space is Borel isomorphic to a measurable subspace of $\mathbb{R}$. We can simply map the Polish space $\mathcal{X}$ to   a measurable subspace $\mathcal{Y}$ of $\mathbb{R}$ via an isomorphism $f$, which naturally   induces a map from  $\mathcal{X}^n$ to  $\mathcal{Y}^n$  via the isomorphism $f^{(n)}:=(f,f,...,f)$. We  then perform the   partition $\{\mathcal{B}_i\}$ for the  high probability subset $\mathcal{A}^c$ of $\mathbb{R}^n$ as above, and take intersections $\{\mathcal{B}_i \cap \mathcal{Y}^n\}$ to produce a partition for the  high probability subset  of $\mathcal{Y}^n$. Mapping this back to the Polish space $\mathcal{X}^n$, we obtain a  desired partition. 

%

Proof of (\ref{eq:shift}): Note that given $q\in[1:n-1]$, we can
divide $[1:n]$ into $k:=\mathrm{gcd}(n,q)$ subsets of size $\frac{n}{k}$,
and the elements in each subset differ by a multiple of $q$ (modulo-$n$).
For example, when $n=6$ and $q=4$, we have two subsets $\{1,3,5\}$
and $\{2,4,6\}$. In this case, $X^{n}=s_{q}^{(n)}(X^{n})$ means
$X_{1}=X_{3}=X_{5}$ and $X_{2}=X_{4}=X_{6}$; as a consequence 
\begin{align*}
\mathbb{P}(X^{n}=s_{q}^{(n)}(X^{n})) & \leq\mathbb{P}(X_{1}=X_{3}\mbox{ and }X_{2}=X_{4})\\
 & =\tau^{2},
\end{align*}
where $\tau:=\mathbb{P}(X_{i}=X_{j})$ for $i\neq j$. In general,
it can be verified that $X^{n}=s_{q}^{(n)}(X^{n})$ implies at least
$\lfloor\frac{n}{2k}\rfloor k$ independent events of the kind $X_{i}\neq X_{j}$,
which, together with the fact $\lfloor\frac{n}{2k}\rfloor k\geq\frac{n}{4}$, further
implies 
\begin{align*}
\mathbb{P}(X^{n}=s_{q}^{(n)}(X^{n}))\leq\tau^{\frac{n}{4}}.
\end{align*}
Now one can readily prove (\ref{eq:shift}) since 
\begin{align*}
\mathbb{P}(X^{n}\in\mathcal{A}) & \leq\sum\limits _{q=1}^{n-1}\mathbb{P}(X^{n}=s_{q}^{(n)}(X^{n}))\\
 & \leq(n-1)\tau^{\frac{n}{4}}\\
 & \rightarrow 0 \mbox{ as }n\rightarrow\infty,
\end{align*}
where the last step is due to $\tau\in[0,1)$.
\end{remark}

\section{Verification of Assumption \ref{assumption:simulation}}\label{app:verification}

Here we verify Assumption \ref{assumption:simulation} for the case $\mathbb{E}[X^2]<\infty$ and $\Delta(x,\hat{x}):=(x-\hat{x})^2$.

According to the definition of $R(D,0)$, for any $D>0$ and $\delta\in(0,D)$, there exists a random variable $\check{X}$ such that
\begin{align*}
&I(X;\check{X})\leq R(D-\delta,0)+\delta,\\
&\mathbb{E}[(X-\check{X})^2]\leq D-\delta,\\
&p_{\check{X}}=p_X.
\end{align*}
Let $\tilde{X}$ be a quantized version of $\check{X}$, obtained by mapping
$\check{X}$ to its nearest point in $\frac{1}{\sqrt{N}}[-N:N]$, where $N$
is a positive integer. By the data processing inequality,
\begin{align*}
I(X;\tilde{X})\leq I(X;\check{X})\leq R(D-\delta,0)+\delta.
\end{align*}
As $R(D,0)$ is convex in $D$ and consequently\footnote{It is shown in Appendix \ref{app:assumptionquantization} that $R(D,0)<\infty$ for $D>0$.} continuous for $D>0$, we have $R(D-\delta,0)+\delta\leq R(D,0)+\epsilon$ by choosing  a sufficiently small $\delta$. 
Let $\hat{X}$ be the mirror version of $\check{X}$ with respect to $\tilde{X}$ in the sense that $p_{X\check{X}\tilde{X}\hat{X}}=p_{X\check{X}}p_{\tilde{X}|\check{X}}p_{\hat{X}|\tilde{X}}$ and $p_{\hat{X}|\tilde{X}}=p_{\check{X}|\tilde{X}}$. Obviously, $X\leftrightarrow \check{X}\leftrightarrow\tilde{X}\leftrightarrow\hat{X}$
form a Markov chain and $p_{\hat{X}}=p_{\check{X}}=p_{X}$. Moreover,
\begin{align}
&\mathbb{E}[(X-\hat{X})^2]\nonumber\\
&=\mathbb{E}[((X-\check{X})+(\check{X}-\tilde{X})+(\tilde{X}-\hat{X}))^2]\nonumber\\
&=\mathbb{E}[(X-\check{X})^2]+\mathbb{E}[(\check{X}-\tilde{X})^2]+\mathbb{E}[(\tilde{X}-\hat{X})^2]\nonumber\\
&\quad+2\mathbb{E}[(X-\check{X})(\check{X}-\tilde{X})]+2\mathbb{E}[(X-\check{X})(\tilde{X}-\hat{X})]\nonumber\\
&\quad+2\mathbb{E}[(\check{X}-\tilde{X})(\tilde{X}-\hat{X})]\nonumber\\
&\leq\mathbb{E}[(X-\check{X})^2]+\mathbb{E}[(\check{X}-\tilde{X})^2]+\mathbb{E}[(\tilde{X}-\hat{X})^2]\nonumber\\
&\quad+2\sqrt{\mathbb{E}[(X-\check{X})^2]\mathbb{E}[(\check{X}-\tilde{X})^2]}\nonumber\\
&\quad+2\sqrt{\mathbb{E}[(X-\check{X})^2]\mathbb{E}[(\tilde{X}-\hat{X})^2]}\nonumber\\
&\quad+2\sqrt{\mathbb{E}[(\check{X}-\tilde{X})^2]\mathbb{E}[(\tilde{X}-\hat{X})^2]}\label{eq:cs}\\
&=\mathbb{E}[(X-\check{X})^2]+4\mathbb{E}[(\check{X}-\tilde{X})^2]\nonumber\\
&\quad+4\sqrt{\mathbb{E}[(X-\check{X})^2]\mathbb{E}[(\check{X}-\tilde{X})^2]}\label{eq:identicaldistribution}\\
&\leq D-\delta+4\mathbb{E}[(\check{X}-\tilde{X})^2]+4\sqrt{(D-\delta)\mathbb{E}[(\check{X}-\tilde{X})^2]},\nonumber
\end{align}
where (\ref{eq:cs}) is due to the Cauchy-Schwarz inequality, and (\ref{eq:identicaldistribution}) is due to $\mathbb{E}[(\check{X}-\tilde{X})^2]=\mathbb{E}[(\tilde{X}-\hat{X})^2]$ (implied by the fact that $p_{\tilde{X}\check{X}}=p_{\tilde{X}\hat{X}}$). Since $p_{\check{X}}=p_{X}$, it follows that $\mathbb{E}[\check{X}^2]<\infty$, which further implies
$\mathbb{E}[(\check{X}-\tilde{X})^2]\rightarrow 0$ as $N\rightarrow\infty$. Therefore, we can ensure $\mathbb{E}[(X-\hat{X})^2]\leq D$ by setting $N$ large enough. This completes the verification.

\section{Proof of Theorem \ref{thm:weakP}}\label{app:weakP} 

According to Assumption \ref{assumption:simulation}, for any $\epsilon>0$, there exist $\tilde{X}$ and $\hat{X}$ such that $X\leftrightarrow \tilde{X}\leftrightarrow \hat{X}$ form a Markov chain, the support of $\tilde{X}$, denoted $\tilde{X}$, satisifies $|\tilde{X}|<\infty$, and
\begin{align*}
&I(X;\tilde{X})\leq R(D,0)+\epsilon,\\
&\mathbb{E}[\Delta(X,\tilde{X})]\leq D,\\
&p_{\hat{X}}=p_{X}.
\end{align*}

We shall treat the conditional distribution $p_{X|\tilde{X}}$ induced by $p_{X\tilde{X}\hat{X}}$ specified above as a memoryless channel and establish a soft-covering lemma that is needed for the proof of Theorem \ref{thm:weakP}. 

\begin{definition}
Given a codebook $\mathcal{C}^{(n)}\subseteq\tilde{\mathcal{X}}^n$, 
let $p^{out}_{\mathcal{C}^{(n)}}$ denote its induced distribution of the output sequence  generated through memoryless channel
$p_{X|\tilde{X}}$ by a codeword randomly picked from $\mathcal{C}^{(n)}$ according to the uniform distribution.
\end{definition}






\begin{lemma}\label{lem:info-spectrum}
	For any $R>I(\tilde{X};X)$ and $\delta>0$, there exists a sequence of codebooks $\{\mathcal{C}^{(n)}\}_{n=1}^\infty$ with $\mathcal{C}^{(n)}\subseteq\mathcal{T}^{(n)}_{\delta}(p_{\tilde{X}})$ and $|\mathcal{C}^{(n)}|\leq 2^{nR}$ such that 
	\begin{align*}
	\lim\limits_{n\rightarrow\infty}d_{\mathrm{TV}}(p^{out}_{\mathcal{C}^{(n)}},p_{X}^{n})=0,
	\end{align*}
	where $\mathcal{T}^{(n)}_{\delta}(p_{\tilde{X}})$ denotes the set of $\delta$-typical sequences with respect to $p_{\tilde{X}}$. 
\end{lemma}

\begin{IEEEproof}
	We first briefly review some basic definitions in information-spectrum methods. 	
	The \emph{limsup in probability} of a sequence of random variables
	$\{W_{n}\}_{n=1}^{\infty}$  is defined
	as 
	\[
	\textrm{p-}\limsup_{n\to\infty}W_{n}:=\inf\left\{ \tau:\lim\limits_{n\rightarrow\infty}\mathbb{P}\{W_{n}>\tau\}=0\right\} .
	\]
	Correspondingly, the \emph{liminf in probability} is defined as 
	\[
	\textrm{p-}\liminf_{n\to\infty}W_{n}:=-\textrm{p-}\limsup_{n\to\infty}-W_{n}.
	\]
	For a sequence of pairs of random variables $(\mathbf{W},\mathbf{V}):=\{ \left(W^{n},V^{n}\right)\} _{n=1}^{\infty}$,
	the \emph{sup-information rate} of $\left(\mathbf{W},\mathbf{V}\right)$
	is defined as 
	\[
	\bar{I}\left(\mathbf{W},\mathbf{V}\right):=\textrm{p-}\limsup_{n\to\infty}\frac{1}{n}\imath_{W^{n};V^{n}}(W^{n};V^{n})
	\]
	where $\imath_{W^{n};V^{n}}:=\log\frac{dp_{W^{n}V^{n}}}{d(p_{W^{n}}p_{V^{n}})}$
	denotes the information density of $(W^{n},V^{n})$. For
	a sequence of distributions $\{ (p_{W^{n}},p_{V^{n}})\} _{n=1}^{\infty}$,
	the \emph{inf-relative-entropy rate} of $\{ (p_{W^{n}},p_{V^{n}})\} _{n=1}^{\infty}$
	is defined as 
	\begin{align*}
	\underline{D}\left(\{p_{W^{n}}\}_{n=1}^{\infty}\|\{p_{V^{n}}\}_{n=1}^{\infty}\right):=\textrm{p-}\liminf_{n\to\infty}\frac{1}{n}\log\frac{d p_{W^{n}}}{dp_{V^{n}}}(W^{n}).
	\end{align*}

Now we are in a position to prove Lemma \ref{lem:info-spectrum}. Construct  a sequence of pairs of random variables $(\tilde{\mathbf{X}},\mathbf{X}):=\{(\tilde{X}^{n},X^{n})\} _{n=1}^{\infty}$ with $(\tilde{X}^{n},X^{n})\sim p^n_{\tilde{X}X}$ and another sequence of pairs of random variables $(\tilde{\mathbf{Y}},\mathbf{Y}):=\{(\tilde{Y}^{n},Y^{n})\} _{n=1}^{\infty}$ with $(\tilde{Y}^{n},Y^{n})\sim p_{\tilde{Y}^n}p_{Y^n|\tilde{Y}^n}$, where $p_{Y^n|\tilde{Y}^n}:=p^n_{X|\tilde{X}}$ and  $p_{\tilde{Y}^n}$ is a truncated version $p^n_{\tilde{X}}$ in the sense that
\begin{align*}
p_{\tilde{Y}^n}(y^n):=\begin{cases}
\frac{p^n_{\tilde{X}}(y^n)}{\mathbb{P}(\tilde{X}^n\in\mathcal{T}^{(n)}_{\delta}(p_{\tilde{X}}))}, & y^n\in\mathcal{T}^n_{\delta}(p_{\tilde{X}}),\\
0 & y^n\in\tilde{X}^n\backslash\mathcal{T}^n_{\delta}(p_{\tilde{X}}).
\end{cases}
\end{align*}
By the general soft-covering lemma \cite[Theorem 4]{Han93}\cite[Corollary VII.4]{Cuff13},
for any $R>\bar{I}(\tilde{\mathbf{Y}};\mathbf{Y})$, there exists a sequence of codebooks with $\mathcal{C}^{(n)}\subseteq\mathcal{T}^{(n)}_{\delta}(p_{\tilde{X}})$ and $|\mathcal{C}^{(n)}|\leq 2^{nR}$ such that 
\begin{align*}
\lim\limits_{n\rightarrow\infty}d_{\mathrm{TV}}(p^{out}_{\mathcal{C}^{(n)}},p_{Y^n})=0.
\end{align*}
Note that
\begin{align}
d_{\mathrm{TV}}(p_{Y^n},p^n_{X})&\leq d_{\mathrm{TV}}(p_{\tilde{Y}^n},p^n_{\tilde{X}})\label{eq:TVdpi}\\
&=\mathbb{P}(\tilde{X}\notin\mathcal{T}^{(\delta)}_n)\nonumber\\
&\rightarrow 0 \mbox{ as }n\rightarrow\infty,\nonumber
\end{align}
where (\ref{eq:TVdpi}) is due to the data processing inequality for the total variation distance. Moreover, in light of the triangle inequality for the total variation distance,
\begin{align*}
d_{\mathrm{TV}}(p^{out}_{\mathcal{C}^{(n)}},p^n_{X})\leq d_{\mathrm{TV}}(p^{out}_{\mathcal{C}^{(n)}},p_{Y^n})+d_{\mathrm{TV}}(p_{Y^n},p^n_{X}).
\end{align*}
So it suffices to prove  $I(\tilde{X};X)\geq\bar{I}(\tilde{\mathbf{Y}};\mathbf{Y})$.

It can be verified that
\begin{align}
\bar{I}(\tilde{\mathbf{Y}};\mathbf{Y}) &=\textrm{p-}\limsup_{n\to\infty}\frac{1}{n}\log\frac{dp_{\tilde{Y}^{n}Y^{n}}}{d\left(p_{\tilde{Y}^{n}}p_{Y^{n}}\right)}(\tilde{Y}^n,Y^n)\nonumber\\
 &=\textrm{p-}\limsup_{n\to\infty}\frac{1}{n}\log\frac{dp_{X|\tilde{X}}^{n}}{dp_{X}^{n}}(\tilde{Y}^{n},Y^{n})\nonumber \\
 & \qquad+\frac{1}{n}\log\frac{dp_{X}^{n}}{d p_{Y^{n}}}(Y^{n})\nonumber \\
 &\leq\textrm{p-}\limsup_{n\to\infty}\frac{1}{n}\log\frac{dp_{X|\tilde{X}}^{n}}{dp_{X}^{n}}(\tilde{Y}^{n},Y^{n})\nonumber \\
 & \qquad+\textrm{p-}\limsup_{n\to\infty}\frac{1}{n}\log\frac{dp_{X}^{n}}{d p_{Y^{n}}}(Y^{n})\label{eq:aplusb} \\
 & =\textrm{p-}\limsup_{n\to\infty}\frac{1}{n}\log\frac{dp_{X|\tilde{X}}^{n}}{dp_{X}^{n}}(\tilde{Y}^{n},Y^{n})\nonumber \\
 & \qquad-\underline{D}(\{p_{Y^{n}}\}_{n=1}^{\infty}\|\{p^n_{X}\}_{n=1}^{\infty}),\nonumber 
\end{align}
where (\ref{eq:aplusb}) follows by
\cite[p. 14, the third inequality from the bottom]{Han03}. Since $\lim_{n\rightarrow\infty}\mathbb{P}(\tilde{X}^n\in\mathcal{T}^{(n)}_{\delta}(p_{\tilde{X}}))=1$ and the conditional distribution of $(\tilde{X}^n,X^n)$ given $\tilde{X}^n\in\mathcal{T}^{(n)}_{\delta}(p_{\tilde{X}})$ is the same as $p_{\tilde{Y}^n,Y^n}$, we must have
\begin{align*}
&\textrm{p-}\limsup_{n\to\infty}\frac{1}{n}\log\frac{dp_{X|\tilde{X}}^{n}}{dp_{X}^{n}}(\tilde{Y}^{n},Y^{n})\\
&=
\textrm{p-}\limsup_{n\to\infty}\frac{1}{n}\log\frac{dp_{X|\tilde{X}}^{n}}{dp_{X}^{n}}(\tilde{X}^{n},X^{n}).
\end{align*}
Moreover, by the weak law of large numbers, 
\begin{align*}
\textrm{p-}\limsup_{n\to\infty}\frac{1}{n}\log\frac{dp_{X|\tilde{X}}^{n}}{dp_{X}^{n}}(\tilde{X}^{n},X^{n})=I(\tilde{X};X).
\end{align*}
Finally, invoking the fact that $\underline{D}(\{p_{Y^{n}}\}_{n=1}^{\infty}\|\{p^n_{X}\}_{n=1}^{\infty})\ge0$ \cite[Lemma 3.2.1]{Han03} completes the proof of Lemma \ref{lem:info-spectrum}.
\end{IEEEproof}

Now we proceed to prove Theorem \ref{thm:weakP}. Given $R>I(\tilde{X};X)$ and $\delta>0$, let $\{\mathcal{C}^{(n)}\}_{n=1}^\infty$ be a sequence of codebooks with the properties specified in Lemma \ref{lem:info-spectrum}. Construct  $\mathcal{C}^{(n)}_q$  by applying the shift operator $s^{(n)}_q$ to every codeword of $\mathcal{C}^{(n)}$, $q\in[0:n-1]$. Let $Q$ be uniformly distributed over $[0:n-1]$, and $\check{Y}^n$ be uniformly distributed over $\mathcal{C}^{(n)}_i$ given $Q=q$, $q\in[0:n-1]$. Denote the output sequence generated by 
$\check{Y}^n$ through memoryless channel $p_{X|\tilde{X}}$ as $\bar{Y}^n$. For $q\in[0:n-1]$,
\begin{align}
d_{\mathrm{TV}}(p_{\bar{Y}^n|Q=q},p^n_X)&=d_{\mathrm{TV}}(p_{\bar{Y}^n|Q=0},p^n_X)\label{eq:shiftinvariant}\\
&=d_{\mathrm{TV}}(p^{out}_{\mathcal{C}^{(n)}},p^n_X),\nonumber
\end{align}
where (\ref{eq:shiftinvariant}) holds because $p_{\bar{Y}^n|Q=q}$ is simply a shifted version of $p_{\bar{Y}^n|Q=0}$ while $p^n_X$ is shift-invariant. Since $d_{\mathrm{TV}}(\cdot,\cdot)$ is convex in its first argument, it follows that
\begin{align*}
d_{\mathrm{TV}}(p_{\bar{Y}^n},p^n_X)\leq d_{\mathrm{TV}}(p^{out}_{\mathcal{C}^{(n)}},p^n_X)
\end{align*}
and consequently must converge to $0$ as $n\rightarrow\infty$.


 
Our coding scheme can be illustrated using the following probablistic graphical model:
\[
\begin{array}{cc}
X^{n} & \searrow\\
\downarrow\\
\check{X}^{n} & \nearrow
\end{array}\tilde{X}^{n}\longrightarrow\hat{X}^{n}
\]
The encoder first leverages the conditional distribution $p_{\check{Y}^n|\bar{Y}^n}$ induced by $p_{\check{Y}^n\bar{Y}^n}$ to generate $\check{X}^n$ based on $X^n$. Note that we have $X^n\in\mathcal{X}^n$, $\check{X}^n\in\cup_{q=0}^{n-1}\mathcal{C}^{(n)}_q$, and $p_{X^n\check{X}^n}=p^n_{X}p_{\check{X}^n|X^n}=p^n_Xp_{\check{Y}^n|\bar{Y}^n}$. 
For $t\in[1:n]$,
\begin{align}
&d_{\mathrm{TV}}(p_{X_t\check{X}_t},p_{X\tilde{X}})\nonumber\\
&\leq d_{\mathrm{TV}}(p_{X_t\check{X}_t},p_{\bar{Y}_t\check{Y}_t})+d_{\mathrm{TV}}(p_{\bar{Y}_t\check{Y}_t},p_{X\tilde{X}})\label{eq:triangleq}\\
&\leq d_{\mathrm{TV}}(p_{X^n\check{X}^n},p_{\bar{Y}^n\check{Y}^n})+d_{\mathrm{TV}}(p_{\bar{Y}_t\check{Y}_t},p_{X\tilde{X}})\label{eq:dpiTV}\\
&=d_{\mathrm{TV}}(p^n_X,p_{\bar{Y}^n})+d_{\mathrm{TV}}(p_{\check{Y}_t},p_{\tilde{X}}),\nonumber
\end{align}
where (\ref{eq:triangleq}) and (\ref{eq:dpiTV}) are due to the triangle inequality and the data processing indequality for the total variation distance, respectively. Furthermore, we have $d_{\mathrm{TV}}(p^n_X,p_{\bar{Y}^n})\rightarrow 0$ as $n\rightarrow\infty$, and $d_{\mathrm{TV}}(p_{\check{Y}_t},p_{\tilde{X}})\rightarrow 0$ uniformly for all $t$ as $\delta\rightarrow 0$. Therefore, 
\begin{align*}
d_{\mathrm{TV}}(p_{X_t\check{X}_t},p_{X\tilde{X}})\leq\epsilon_{\delta}
\end{align*}
for $t\in[1:n]$ and all sufficiently large $n$, where $\epsilon_{\delta}\rightarrow 0$ as $\delta\rightarrow 0$.

For each $t\in[1:n]$, let\footnote{Since $p_{\check{Y}^n}$ is shift-invarint and $p_{X|\tilde{X}}$ is memoryless, it follows that $p_{\bar{Y}^n\check{Y}^n}$ and consequently $p_{\check{Y}^n|\bar{Y}^n}$ are shift-invariant as well, which further implies the shift-invariance of $p_{X^n\check{X}^n}$ in view of the fact $p_{X^n\check{X}^n}=p^n_Xp_{\check{Y}^n|\bar{Y}^n}$. Therefore, $p_{\check{X}_t|X_t}$, $b_{X'_t|X_t}$, $r_{\check{X}_t|X_t}$, $r_{\tilde{X}_t|X_t}$, $\kappa_t$, and $p_{\check{X}_t\tilde{X}_t|X_t}$ are all time-invariant.}
	\begin{align*}
	&p_{\check{X}_{t}\tilde{X}_{t}|X_{t}}(\check{x},\tilde{x}|x)\\
	&:=b_{X'_t|X_t}(\check{x}|x)1_{\check{x}=\tilde{x}}({\check{x},\tilde{x}})\\
	&\quad+
	\begin{cases}
	\frac{1}{\kappa_t(x)}r_{\check{X}_t|X_t}(\check{x}|x)r_{\tilde{X}_t|X_t}(\tilde{x}|x),& \kappa_t(x)>0,\\
	0, & \kappa_t(x)=0,
	\end{cases}\\
	&\hspace{1.5in}(x,\check{x},\tilde{x})\in\mathcal{X}\times\tilde{\mathcal{X}}\times\tilde{\mathcal{X}},
	\end{align*}
	where 
		\begin{align*}
	b_{X'_t|X_t}(x'|x) &:=\min\{p_{\check{X}_{t}|X_{t}}(x'|x),p_{\tilde{X}|X}(x'|x)\},\\
	r_{\check{X}_t|X_t}(\check{x}|x) & :=p_{\check{X}_{t}|X_{t}}(\check{x}|x)-b_{X'_t|X_t}(\check{x}|x),\\
	r_{\tilde{X}_t|X_t}(\tilde{x}|x) & :=p_{\tilde{X}|X}(\tilde{x}|x)-b_{X'_t|X_t}(\tilde{x}|x),\\
	\kappa_t(x) & :=d_{\mathrm{TV}}(p_{\check{X}_{t}|X_{t}=x},p_{\tilde{X}|X=x})\\
	& =\sum_{\check{x}\in\tilde{\mathcal{X}}}r_{\check{X}_t|X_t}(\check{x}|x)\\
	& =\sum_{\tilde{x}\in\tilde{\mathcal{X}}}r_{\tilde{X}_t|X_t}(\tilde{x}|x)\\
	& =1-\sum_{x'\in\tilde{\mathcal{X}}}b_{X'_t|X_t}(x'|x).
	\end{align*}
	We claim that $p_{\check{X}_{t}\tilde{X}_{t}|X_{t}}$ is
	 a regular conditional distribution. Indeed, 
	on one hand, given each $x$, $p_{\check{X}_{t}\tilde{X}_{t}|X_{t}=x}$
	is a distribution (more precisely, a probability mass function) since $p_{\check{X}_{t}\tilde{X}_{t}|X_{t}}(\check{x},\tilde{x}|x)\geq 0$, $(\check{x},\tilde{x})\in\tilde{\mathcal{X}}^2$, and
	\begin{align*}
	\sum\limits_{\check{x},\tilde{x}\in\tilde{\mathcal{X}}}p_{\check{X}_{t}\tilde{X}_{t}|X_{t}}(\check{x},\tilde{x}|x)&=\sum\limits_{\check{x}}b_{X'_t|X_t}(\check{x}|x)+\kappa_t(x)=1;
	\end{align*}
	on the other hand, given each $(\check{x},\tilde{x})$, $x\mapsto p_{\check{X}_{t}\tilde{X}_{t}|X_{t}}(\check{x},\tilde{x}|x)$
	is measurable (which is due to the fact that both $x\mapsto p_{\check{X}_{t}|X_{t}}(\check{x}|x)$ and $x\mapsto p_{\tilde{X}|X}(\tilde{x}|x)$
	are measurable), and so is $x\mapsto\check{p}_{\check{X}_{t}\tilde{X}_{t}|X_{t}}(\mathcal{B}|x)$
	for each $\mathcal{B}\subseteq\tilde{\mathcal{X}}^{2}$. Moreover, given each $x$, $\check{p}_{\check{X}_{t}\tilde{X}_{t}|X_{t}=x}$
	is in fact a maximal coupling of $p_{\check{X}_{t}|X_{t}=x}$ and
	$p_{\tilde{X}|X=x}$ since
	\begin{align*}
	&\sum\limits_{\check{x},\tilde{x}\in\tilde{\mathcal{X}}:\check{x}\neq\tilde{x}}p_{\check{X}_t\tilde{X}_t|X_t}(\check{x},\tilde{x}|x)\\
	&=\frac{1}{\kappa_t(x)}\sum\limits_{\check{x}\in\tilde{\mathcal{X}}}r_{\check{X}_t|X_t}(\check{x}|x)\sum\limits_{\tilde{x}\in\tilde{\mathcal{X}}:\tilde{x}\neq\check{x}}r_{\tilde{X}_t|X_t}(\tilde{x}|x)\\
	&=\frac{1}{\kappa_t(x)}\sum\limits_{\check{x}\in\tilde{\mathcal{X}}}r_{\check{X}_t|X_t}(\check{x}|x)(\kappa_t(x)-r_{\tilde{X}_t|X_t}(\check{x}|x))\\
	&=\kappa_t(x)-\frac{1}{\kappa_t(x)}\sum\limits_{\check{x}\in\tilde{\mathcal{X}}}r_{\check{X}_t|X_t}(\check{x}|x)r_{\tilde{X}_t|X_t}(\check{x}|x)\\
	&=\kappa_t(x)\\
	&=d_{\mathrm{TV}}(\check{p}_{\tilde{X}_{i}|X_{i}=x}, p_{\tilde{X}|X=x}),\quad\kappa_t(x)>0,
	\end{align*}
	which clearly also holds when $\kappa_t(x)=0$.
	



The encoder leverages $p_{\tilde{X}_t|X_t\check{X}_t}$ induced by $p_{X_t\check{X}_t\tilde{X}_t}:=p_Xp_{\check{X}_t\tilde{X}_t|X_t}$ to generate $\tilde{X}_t$ from $(X_t,\check{X}_t)$, $t\in[1:n]$. Note that $p_{X_t\tilde{X}_t}=p_{X\tilde{X}}$ and 
\begin{align*}
\mathbb{P}(\check{X}_{t}\neq\tilde{X}_{t}) & =\int\mathbb{P}(\check{X}_{t}\neq\tilde{X}_{t}|X_{t}=t)dp_{X}(x)\\
& =\int d_{\mathrm{TV}}(p_{\check{X}_{t}|X_{t}=x},p_{\tilde{X}|X=x})dp_{X}(x)\\
& =d_{\mathrm{TV}}(p_{X_t\check{X}_t}, p_{X\tilde{X}})\\
& \leq\epsilon_{\delta},\quad t\in[1:n],
\end{align*}
when $n$ is sufficiently large.


The encoder then sends $\tilde{X}^n$ to the decoder. We have
\begin{align}
\frac{1}{n}H(\tilde{X}^n)&\leq\frac{1}{n}H(\check{X}^n)+\frac{1}{n}H(\tilde{X}^n|\check{X}^n)\nonumber\\
&\leq R+\frac{\log n}{n}+\frac{1}{n}H(\tilde{X}^n|\check{X}^n)\nonumber\\
&\leq R+\frac{\log n}{n}+\frac{1}{n}\sum\limits_{t=1}^nH(\tilde{X}_t|\check{X}_t)\nonumber\\
&\leq R+\frac{\log n}{n}+\frac{1}{n}\sum\limits_{t=1}^n (H_b(\mathbb{P}(\check{X}_t\neq\tilde{X}_t))\nonumber\\
&\hspace{1.3in}+\mathbb{P}(\check{X}_t\neq\tilde{X}_t)\log|\tilde{\mathcal{X}}|)\label{eq:Fano}\\
&\leq R+\frac{\log n}{n}+H_b(\epsilon_{\delta})+\epsilon_{\delta}\log|\tilde{\mathcal{X}}|,\label{eq:half}
\end{align}
where (\ref{eq:Fano}) is due to Fano's inequality, and (\ref{eq:half}) holds when $\epsilon_{\delta}\leq\frac{1}{2}$ and $n$ is sufficiently large.


Given $\tilde{X}^n$, the decoder simply generates $\hat{X}^n$ using the conditional distribution $p_{\hat{X}^n|\tilde{X}^n}:=p^n_{\hat{X}|\tilde{X}}$. 
It is clear that $p_{X_t\hat{X}_t}=p_{X\hat{X}}$, $t\in[1:n]$, and consequently
\begin{align*}
& \frac{1}{n}\sum\limits _{t=1}^{n}\mathbb{E}[\Delta(X_{t},\hat{X}_{t})]=\mathbb{E}[\Delta(X,\hat{X})]\leq D,\\
& \check{p}_{\hat{X}_{t}}=p_{X},\quad t\in[1:n].
\end{align*}
Finally, by choosing $\epsilon, \delta$ sufficiently small, $R$ sufficiently close to $I(X;\tilde{X})$, and $n$ sufficiently large, we can make
\begin{align*}
 R+\frac{\log n}{n}+H_b(\epsilon_{\delta})+\epsilon_{\delta}\log|\tilde{\mathcal{X}}|
\end{align*}
as close to $R(D,0)$ as we want. 
This completes the proof of Theorem \ref{thm:weakP}.




\section{Verification of Uniform Integrability}\label{app:uniformintegrability}

Here we verify uniform integrability for the case $\mathbb{E}[X^2]<\infty$, $\mathbb{E}[\hat{X}^2]<\infty$, and $\Delta(x,\hat{x}):=(x-\hat{x})^2$.

By the Cauchy-Schwarz inequality,
\begin{align*}
&\mathbb{E}[(X-\hat{X})^21_{\mathcal{E}}(X,\hat{X})]\\
&\leq(\sqrt{\mathbb{E}[X^21_{\mathcal{E}}(X,\hat{X})]}+\sqrt{\mathbb{E}[\hat{X}^21_{\mathcal{E}}(X,\hat{X})]})^2.
\end{align*}
Note that for any $\chi>0$,
\begin{align*}
&\mathbb{E}[X^21_{\mathcal{E}}(X,\hat{X})]\\
&=\mathbb{P}(X^2\leq \chi)\mathbb{E}[X^21_{\mathcal{E}}(X,\hat{X})|X^2\leq \chi]\\
&\quad+\mathbb{P}(X^2> \chi)\mathbb{E}[X^21_{\mathcal{E}}(X,\hat{X})|X^2>\chi]\\
&\leq \chi\mathbb{P}((X,\hat{X})\in\mathcal{E})+\mathbb{P}(X^2> \chi)\mathbb{E}[X^2|X^2> \chi]\\
&\leq \chi\delta+\mathbb{P}(X^2> \chi)\mathbb{E}[X^2|X^2> \chi].
\end{align*}
Similarly, we have
\begin{align*}
\mathbb{E}[\hat{X}^21_{\mathcal{E}}(X,\hat{X})]\leq \chi\delta+\mathbb{P}(\hat{X}^2> \chi)\mathbb{E}[\hat{X}^2|\hat{X}^2> \chi].
\end{align*}
Since $\mathbb{E}[X^2]<\infty$ and $\mathbb{E}[\hat{X}^2]\}<\infty$, it follows by the dominated convergence theorem that both $\mathbb{P}(X^2> \chi)\mathbb{E}[X^2|X^2> \chi]$ and $\mathbb{P}(\hat{X}^2> \chi)\mathbb{E}[\hat{X}^2|\hat{X}^2>\chi]$
converge to $0$ as $\chi\rightarrow\infty$. Therefore, there exists $\chi^*>0$ such that
$\mathbb{P}(X^2> \chi^*)\mathbb{E}[X^2|X^2> \chi^*]\leq\frac{\epsilon}{8}$ and $\mathbb{P}(\hat{X}^2> \chi^*)\mathbb{E}[\hat{X}^2|\hat{X}^2> \chi^*]\leq\frac{\epsilon}{8}$. Setting $\delta=\frac{\epsilon}{8\chi^*}$ ensures
\begin{align*}
\mathbb{E}[(X-\hat{X})^21_{\mathcal{E}}(X,\hat{X})]\leq\left(2\sqrt{\chi^*\frac{\epsilon}{8\chi^*}+\frac{\epsilon}{8}}\right)^2=\epsilon.
\end{align*}

\section{Proof of Theorem \ref{thm:prstrong}}

\label{app:prstrong}

We shall first prove $\varphi(D)\geq R(\frac{D}{2})$. For any $p_{U\hat{X}|X}$
satisfying (\ref{eq:constraint1}), (\ref{eq:constraint2}), and (\ref{eq:constraint3}),
let $V:=\mathbb{E}[X|U]$ and $\hat{V}:=\mathbb{E}[\hat{X}|U]$. Since
$X\leftrightarrow U\leftrightarrow\hat{X}$ form a Markov chain (see
(\ref{eq:constraint2})), it follows that 
\begin{align*}
 & \mathbb{E}[(X-\hat{X})^{2}]\\
 & =\mathbb{E}[(X-V)^{2}]+\mathbb{E}[(V-\hat{V})^{2}]+\mathbb{E}[(\hat{X}-\hat{V})^{2}],
\end{align*}
which, together with (\ref{eq:constraint1}), implies 
\begin{align*}
\min\{\mathbb{E}[(X-V)^{2}],\mathbb{E}[(\hat{X}-\hat{V})^{2}]\}\leq\frac{D}{2}.
\end{align*}
Now consider the case $\mathbb{E}[(X-V)^{2}]\leq\frac{D}{2}$. Note
that 
\begin{align}
\max\{I(X;U);I(\hat{X};U)\} & \geq I(X;U)\nonumber \\
 & \geq I(X;V)\label{eq:dpi}\\
 & \geq R(\frac{D}{2}),\nonumber 
\end{align}
where (\ref{eq:dpi}) is due to the data processing inequality. By
symmetry (see (\ref{eq:constraint3})), $\max\{I(X;U);I(\hat{X};U)\}\geq R(\frac{D}{2})$
continues to hold if $\mathbb{E}[(\hat{X}-\hat{V})^{2}]\leq\frac{D}{2}$.
This proves $\varphi(D)\geq R(\frac{D}{2})$.

Next we proceed to prove $\varphi(D)\leq R(\frac{D}{2})$. For any
$p_{V|X}$ satisfying (\ref{eq:constraintV}), let $U:=\mathbb{E}[X|V]$.
We have 
\begin{align*}
\mathbb{E}[(X-U)^{2}]\leq\mathbb{E}[(X-V)^{2}],
\end{align*}
which, together with (\ref{eq:constraintV}), implies $\mathbb{E}[(X-U)^{2}]\leq\frac{D}{2}$.
Now construct $p_{\hat{X}U|X}$ such that $p_{\hat{X}U|X}=p_{U|X}p_{\hat{X}|U}$
and $p_{\hat{X}|U}=p_{X|U}$. Note that (\ref{eq:constraint2}) and
(\ref{eq:constraint3}) are satisfied. Moreover, 
\begin{align*}
\mathbb{E}[(X-\hat{X})^{2}] & =\mathbb{E}[(X-U)^{2}]+\mathbb{E}[(\hat{X}-U)^{2}]\\
 & =2\mathbb{E}[(X-U)^{2}]\\
 & \leq D.
\end{align*}
So (\ref{eq:constraint1}) is also satisfied. As a consequence, 
\begin{align*}
\varphi(D)\leq\max\{I(X;U),I(\hat{X};U)\}.
\end{align*}
The proof is complete in view of the fact that 
\begin{align*}
\max\{I(X;U),I(\hat{X};U)\}=I(X;U)\leq I(X;V).
\end{align*}

\section{Proof of Theorem \ref{thm:binary}}

\label{app:binary}

In view of (\ref{eq:double}), the problem boils down to determining
$R(\frac{D}{2})$ by solving the optimization problem in (\ref{eq:optimization}).
To this end, we need the following lemma. \begin{lemma}\label{lem:KKT}
If there exist $p_{V}^{*}$ over a finite set $\mathcal{V}\subseteq[0,1]$
with $p_{V}^{*}(v)>0$, $v\in\mathcal{V}$, and $\lambda\geq0$ such
that 
\begin{align}
 & \frac{(1-\rho)2^{-\lambda v^{2}}}{\sum_{\tilde{v}\in\mathcal{V}}p_{V}^{*}(\tilde{v})2^{-\lambda\tilde{v}^{2}}}+\frac{\rho2^{-\lambda(1-v)^{2}}}{\sum_{\tilde{v}\in\mathcal{V}}p_{V}^{*}(\tilde{v})2^{-\lambda(1-\tilde{v})^{2}}}=1,\nonumber \\
 & \hspace{2.1in}v\in\mathcal{V},\label{eq:KKT1}\\
 & \frac{(1-\rho)2^{-\lambda v^{2}}}{\sum_{\tilde{v}\in\mathcal{V}}p_{V}^{*}(\tilde{v})2^{-\lambda\tilde{v}^{2}}}+\frac{\rho2^{-\lambda(1-v)^{2}}}{\sum_{\tilde{v}\in\mathcal{V}}p_{V}^{*}(\tilde{v})2^{-\lambda(1-\tilde{v})^{2}}}\leq1,\nonumber \\
 & \hspace{2.1in}v\in[0,1]\backslash\mathcal{V},\label{eq:KKT2}\\
 & (1-\rho)\frac{\sum_{v\in\mathcal{V}}p_{V}^{*}(v)2^{-\lambda v^{2}}v^{2}}{\sum_{\tilde{v}\in\mathcal{V}}p_{V}^{*}(\tilde{v})2^{-\lambda\tilde{v}^{2}}}\nonumber \\
 & +\rho\frac{\sum_{v\in\mathcal{V}}p_{V}^{*}(v)2^{-\lambda(1-v)^{2}}(1-v)^{2}}{\sum_{\tilde{v}\in\mathcal{V}}p_{V}^{*}(\tilde{v})2^{-\lambda(1-\tilde{v})^{2}}}=\frac{D}{2},\label{eq:KKT3}
\end{align}
then $p_{V|X}^{*}$ given by 
\begin{align*}
 & p_{V|X}^{*}(v|x):=\frac{p_{V}^{*}(v)2^{-\lambda(x-v)^{2}}}{\sum_{\tilde{v}\in\mathcal{V}}p_{V}^{*}(\tilde{v})2^{-\lambda(x-\tilde{v})^{2}}},\\
 & \hspace{1in}x\in\{0,1\},\quad v\in\mathcal{V},
\end{align*}
is an optimal solution to (\ref{eq:optimization}). \end{lemma} 
\begin{IEEEproof}
It is clear that there is no loss of generality in assuming that $V$
only takes value from $[0,1]$. Let $\mathcal{V}'$ be an arbitrary
finite subset of $[0,1]$. In view of the standard Karush-Kuhn-Tucker
conditions \cite[pp. 362--364]{CT91}, (\ref{eq:KKT1})--(\ref{eq:KKT3})
ensures that $p_{V|X}^{*}$ attains the infimum in (\ref{eq:optimization})
when the alphabet of $V$ is restricted\footnote{We set $p_{V|X}^{*}(v|x)=0$ for $v\in\mathcal{V}'\backslash\mathcal{V}$.}
to be $\mathcal{V}\cup\mathcal{V}'$. Moreover, according to the support
lemma \cite[p. 631]{EGK11}, it suffices to consider the finite alphabet
case; in fact, the alphabet size of $V$ does not need to exceed $3$
for the purpose of preserving $p_{X}$, $H(X|V)$, and $\mathbb{E}[(X-V)^{2}]$.
So $p_{V|X}^{*}$ must be an optimal solution to (\ref{eq:optimization}). 
\end{IEEEproof}
Now we are in a position to solve (\ref{eq:optimization}). It suffices
to consider the case $D\in(0,2\rho(1-\rho))$ since obviously $R(\frac{D}{2})$
equals $H_{b}(\rho)$ when $D=0$ and equals $0$ when $D\geq2\rho(1-\rho)$.

Let $\mathcal{V}:=\{a,1-a\}$ with 
\begin{align*}
a:=\frac{1-\sqrt{1-2D}}{2}.
\end{align*}
Note that $a\in(0,\rho)$. Define $p_{V}^{*}$ over $\mathcal{V}$
such that 
\begin{align*}
p_{V}^{*}(a)=\frac{1-a-\rho}{1-2a},\quad p_{V}^{*}(1-a)=\frac{\rho-a}{1-2a}.
\end{align*}
Moreover, let 
\begin{align*}
\lambda:=\frac{1}{1-2a}\log(\frac{1-a}{a}),
\end{align*}
which is clearly positive. We shall proceed to verify that the constructed
$p_{V}^{*}$ and $\lambda$ satisfying (\ref{eq:KKT1})--(\ref{eq:KKT3}).

Note that 
\begin{align}
 & \frac{(1-\rho)2^{-\lambda a^{2}}}{p_{V}^{*}(a)2^{-\lambda a^{2}}+p_{V}^{*}(1-a)2^{-\lambda(1-a)^{2}}}\nonumber \\
 & +\frac{\rho2^{-\lambda(1-a)^{2}}}{p_{V}^{*}(a)2^{-\lambda(1-a)^{2}}+p_{V}^{*}(1-a)2^{-\lambda a^{2}}}\nonumber \\
 & =\frac{(1-\rho)}{p_{V}^{*}(a)+p_{V}^{*}(1-a)\frac{a}{1-a}}+\frac{\rho\frac{a}{1-a}}{p_{V}^{*}(a)\frac{a}{1-a}+p_{V}^{*}(1-a)}\label{eq:prop}\\
 & =1,\nonumber 
\end{align}
where (\ref{eq:prop}) is due to 
\begin{align*}
2^{-\lambda(1-a)^{2}}=\frac{a}{1-a}2^{-\lambda a^{2}}.
\end{align*}
Similarly, 
\begin{align*}
 & \frac{(1-\rho)2^{-\lambda(1-a)^{2}}}{p_{V}^{*}(a)2^{-\lambda a^{2}}+p_{V}^{*}(1-a)2^{-\lambda(1-a)^{2}}}\\
 & +\frac{\rho2^{-\lambda a^{2}}}{p_{V}^{*}(a)2^{-\lambda(1-a)^{2}}+p_{V}^{*}(1-a)2^{-\lambda a^{2}}}\\
 & =\frac{(1-\rho)\frac{a}{1-a}}{p_{V}^{*}(a)+p_{V}^{*}(1-a)\frac{a}{1-a}}+\frac{\rho}{p_{V}^{*}(a)\frac{a}{1-a}+p_{V}^{*}(1-a)}\\
 & =1.
\end{align*}
So (\ref{eq:KKT1}) indeed holds.

Next let 
\begin{align*}
\eta(v) & :=\frac{(1-\rho)2^{-\lambda v^{2}}}{p_{V}^{*}(a)2^{-\lambda a^{2}}+p_{V}^{*}(1-a)2^{-\lambda(1-a)^{2}}}\\
 & \quad+\frac{\rho2^{-\lambda(1-v)^{2}}}{p_{V}^{*}(a)2^{-\lambda(1-a)^{2}}+p_{V}^{*}(1-a)2^{-\lambda a^{2}}}.
\end{align*}
We have 
\begin{align*}
\frac{{\rm {d}}}{{\rm d}v}\eta(v) & =-\frac{\frac{2}{\log e}(1-\rho)\lambda v2^{-\lambda v^{2}}}{p_{V}^{*}(a)2^{-\lambda a^{2}}+p_{V}^{*}(1-a)2^{-\lambda(1-a)^{2}}}\\
 & \quad+\frac{\frac{2}{\log2}\rho\lambda(1-v)2^{-\lambda(1-v)^{2}}}{p_{V}^{*}(a)2^{-\lambda(1-a)^{2}}+p_{V}^{*}(1-a)2^{-\lambda a^{2}}}.
\end{align*}
Clearly, 
\begin{align*}
\frac{{\rm {d}}}{{\rm d}v}\eta(v)\mbox{ }\substack{>\\
=\\
<
}
\mbox{ }0
\end{align*}
if and only if 
\begin{align*}
\xi(v)\mbox{ }\substack{>\\
=\\
<
}
\mbox{ }\log\left(\frac{p_{V}^{*}(a)2^{-\lambda(1-a)^{2}}+p_{V}^{*}(1-a)2^{-\lambda a^{2}}}{p_{V}^{*}(a)2^{-\lambda a^{2}}+p_{V}^{*}(1-a)2^{-\lambda(1-a)^{2}}}\right),
\end{align*}
where 
\begin{align*}
\xi(v):=\log\left(\frac{\rho(1-v)2^{-\lambda(1-v)^{2}}}{(1-\rho)v2^{-\lambda v^{2}}}\right).
\end{align*}
It can be verified that 
\begin{align*}
 & \log\left(\frac{p_{V}^{*}(a)2^{-\lambda(1-a)^{2}}+p_{V}^{*}(1-a)2^{-\lambda a^{2}}}{p_{V}^{*}(a)2^{-\lambda a^{2}}+p_{V}^{*}(1-a)2^{-\lambda(1-a)^{2}}}\right)\\
 & =\log\left(\frac{p_{V}^{*}(a)\frac{a}{1-a}+p_{V}^{*}(1-a)}{p_{V}^{*}(a)+p_{V}^{*}(1-a)\frac{a}{1-a}}\right)\\
 & =\log(\frac{\rho}{1-\rho}).
\end{align*}
On the other hand, 
\begin{align*}
\left.\xi(v)\right|_{v=a,\frac{1}{2},1-a}=\log(\frac{\rho}{1-\rho}).
\end{align*}
Moreover, 
\begin{align*}
\frac{{\rm {d}^{2}}}{{\rm d}v^{2}}\xi(v)=\frac{(1-2v)}{v^{2}(1-v)^{2}}\log e,
\end{align*}
which shows that $\xi(v)$ is a strictly convex function for $v\in(0,\frac{1}{2})$
and a strictly concave function for $v\in(\frac{1}{2},1)$. So we
must have 
\begin{align*}
\xi(v)\begin{cases}
>\log(\frac{\rho}{1-\rho}), & v\in[0,a)\cup(\frac{1}{2},1-a),\\
=\log(\frac{\rho}{1-\rho}), & v=a,\frac{1}{2},1-a,\\
<\log(\frac{\rho}{1-\rho}), & v\in(a,\frac{1}{2})\cup(1-a,1],
\end{cases}
\end{align*}
and consequently 
\begin{align*}
\frac{{\rm {d}}}{{\rm d}v}\eta(v)\begin{cases}
>0, & v\in[0,a)\cup(\frac{1}{2},1-a),\\
=0, & v=a,\frac{1}{2},1-a,\\
<0, & v\in(a,\frac{1}{2})\cup(1-a,1].
\end{cases}
\end{align*}
This together with (\ref{eq:KKT1}) implies (\ref{eq:KKT2}).

Finally, we have 
\begin{align*}
 & (1-\rho)\frac{p_{V}^{*}(a)2^{-\lambda a^{2}}a^{2}+p_{V}^{*}(1-a)2^{-\lambda(1-a)^{2}}(1-a)^{2}}{p_{V}^{*}(a)2^{-\lambda a^{2}}+p_{V}^{*}(1-a)2^{-\lambda(1-a)^{2}}}\\
 & +\rho\frac{p_{V}^{*}(a)2^{-\lambda(1-a)^{2}}(1-a)^{2}+p_{V}^{*}(1-a)2^{-\lambda a^{2}}a^{2}}{p_{V}^{*}(a)2^{-\lambda(1-a)^{2}}+p_{V}^{*}(1-a)2^{-\lambda a^{2}}}\\
 & =(1-\rho)\frac{p_{V}^{*}(a)a^{2}+p_{V}^{*}(1-a)a(1-a)}{p_{V}^{*}(a)+p_{V}^{*}(1-a)\frac{a}{1-a}}\\
 & \quad+\rho\frac{p_{V}^{*}(a)a(1-a)+p_{V}^{*}(1-a)a^{2}}{p_{V}^{*}(a)\frac{a}{1-a}+p_{V}^{*}(1-a)}\\
 & =\frac{D}{2},
\end{align*}
which verifies (\ref{eq:KKT3}).

In light of Lemma \ref{lem:KKT}, $p_{V|X}^{*}$ is an optimal solution
to (\ref{eq:optimization}). Note that 
\begin{align*}
p_{V|X}^{*}(a|0) & =\frac{p_{V}^{*}(a)2^{-\lambda a^{2}}}{p_{V}^{*}(a)2^{-\lambda a^{2}}+p_{V}^{*}(1-a)2^{-\lambda(1-a)^{2}}}\\
 & =\frac{p_{V}^{*}(a)}{p_{V}^{*}(a)+p_{V}^{*}(1-a)\frac{a}{1-a}}\\
 & =\frac{(1-a)(1-a-\rho)}{(1-\rho)(1-2a)},\\
p_{V|X}^{*}(a|1) & =\frac{p_{V}^{*}(a)2^{-\lambda(1-a)^{2}}}{p_{V}^{*}(a)2^{-\lambda(1-a)^{2}}+p_{V}^{*}(1-a)2^{-\lambda a^{2}}}\\
 & =\frac{p_{V}^{*}(a)\frac{a}{1-a}}{p_{V}^{*}(a)\frac{a}{1-a}+p_{V}^{*}(1-a)}\\
 & =\frac{a(1-a-\rho)}{\rho(1-2a)},
\end{align*}
and 
\begin{align*}
p_{V|X}^{*}(1-a|0) & =\frac{a(\rho-a)}{(1-\rho)(1-2a)},\\
p_{V|X}^{*}(1-a|1) & =\frac{(1-a)(\rho-a)}{\rho(1-2a)}.
\end{align*}
The induced $p_{X|V}^{*}$ is given by 
\begin{align*}
 & p_{X|V}^{*}(0|a)=1-a,\quad p_{V|X}^{*}(1|a)=a,\\
 & p_{X|V}^{*}(0|1-a)=a,\quad p_{V|X}^{*}(1|1-a)=1-a.
\end{align*}
As a consequence, 
\begin{align*}
R(\frac{D}{2}) & =H_{b}(\rho)-H_{b}(a)\\
 & =H_{b}(\rho)-H_{b}(\frac{1-\sqrt{1-2D}}{2}).
\end{align*}


\end{document}